\documentclass[prl,twocolumn,aps,superscriptaddress,amsmath,amssymb,10pt]{revtex4-1}
\usepackage[utf8]{inputenc}
\usepackage[T1]{fontenc}

\usepackage{graphicx}
\usepackage{framed}
\usepackage{esvect}
\usepackage{amsmath,amssymb,color,bm}
\usepackage{mathrsfs}
\usepackage[dvipsnames]{xcolor}
\usepackage{siunitx}
\usepackage{hyperref}
\hypersetup{
	colorlinks=true,
	linkcolor=blue,
	citecolor=blue,
	filecolor=blue,
	urlcolor=blue,
	pdfstartview=FitH,
	pdfpagemode=UseNone
}

\newcommand\tb[1]{\textbf{#1}}

\newcommand\mc[1]{\mathcal{#1}}

\newcommand\im[1]{\textnormal{Im}\left[#1\right]}
\newcommand\e[1]{\cdot 10^{#1}}

\newcommand\beq{\begin{equation}}
\newcommand\eeq{\end{equation}}
\newcommand\beqa{\begin{eqnarray}}
\newcommand\eeqa{\end{eqnarray}}
 
  \newcommand{\rev}[1]{\textcolor{black}{#1}}

\def\exp{\textnormal{exp}}

\renewcommand{\d}{\mathrm{d}}
\renewcommand{\r}{\mathbf{r}}
\def\q{\tb{q}}

\def\w{\omega}

\begin{document}

\title{Fluctuation-Induced Supersolidity at the Superfluid-Solid Interface }

\author{Baptiste Coquinot}\email{Baptiste.Coquinot@ist.ac.at}
\affiliation{Institute of Science and Technology Austria (ISTA), Am Campus 1, 3400 Klosterneuburg, Austria}

\author{Ragheed Alhyder}
\affiliation{Institute of Science and Technology Austria (ISTA), Am Campus 1, 3400 Klosterneuburg, Austria}

\author{Alberto Cappellaro}
\affiliation{Institute of Science and Technology Austria (ISTA), Am Campus 1, 3400 Klosterneuburg, Austria}

\author{Mikhail Lemeshko}
\email{Mikhail.Lemeshko@ista.ac.at}
\affiliation{Institute of Science and Technology Austria (ISTA), Am Campus 1, 3400 Klosterneuburg, Austria}

\date{\today}

\begin{abstract}
Supersolidity, combining superfluid and crystalline orders, has been realized in dipolar Bose-Einstein condensates by tuning interatomic interactions.
Here we show that supersolidity can also emerge from mode coupling at a superfluid-solid interface, without modifying bulk interactions and for a broad class of superfluids.
Using an analytical and numerical treatment of the coupled superfluid and phonon fields, we derive the criterion for a density-modulation instability driven by interfacial coupling and dependent on dimensionality.
{In superfluid helium, the instability first appears at the roton mode, while in a Bose-Einstein condensate with contact interactions it can occur for a large range of wavevectors.}
Beyond the threshold, the ground state acquires an interfacial density modulation while the bulk remains superfluid, forming a hybrid superfluid-supersolid phase.
{Our results identify interfacial mode coupling as a promising route to supersolidity, enabling the simultaneous exploitation of interfacial supersolid and bulk superfluid quantum properties, as well as using the interface as a seed to nucleate bulk supersolidity. }
\end{abstract}

\maketitle


Superfluidity and supersolidity are two macroscopic manifestations of quantum mechanics.
Superfluidity arises when the $U(1)$ phase symmetry of the wavefunction is spontaneously broken, allowing the fluid to sustain frictionless flow associated with a macroscopic phase gradient.
Bosonic superfluids are now well established experimentally, both in liquid helium and in Bose-Einstein condensates (BECs)~\cite{Griffin1996, Dalfovo1999, Glyde2017, Halperin2018}. 
Extending this concept, a supersolid combines superfluid and crystalline order by further breaking continuous translational symmetry of the density, thus forming a lattice~\cite{Thouless1969, Leggett1970, Chester1970}.
{In helium, the existence of supersolidity remains controversial~\cite{Wessel2005, Chan2008, Balibar2010, Cazorla2010, Silvestrelli2019}. 
In particular, close to solid walls, a strong and uniform absorption leads to the formation of a triangular lattice in the first layer, which is no longer superfluid, while the next layers present supersolid features~\cite{Crowell1993, Boninsegni2011, Nyeki2017, Nyeki2017a, Knapp2025}.
In BECs, supersolidity has also been investigated~\cite{Wang2010, Henkel2010, Datta2010, Anderson2012, Li2012,  Henkel2010, Biagioni2022, Zhang2023}
 and was recently achieved experimentally~\cite{Leonard2017, Li2017, Schuster2020, Liebster2025}, in particular with dipolar BECs~\cite{Chomaz2019, Bottcher2019, Tanzi2019, Tanzi2019a, Sohmen2021, Alana2023, Recati2023, Bigagli2024}.}
 In these last systems, a finite-wavevector roton softening produces a density modulation that remains phase coherent, stabilized by quantum fluctuations. 
 Here, guided by this mechanism, we propose that mode coupling at a superfluid-solid interface hybridizes the superfluid quasiparticles with surface phonons of the solid and induces a softening without delicate interaction tuning.
 For strong enough superfluid-solid coupling, we show that this softening leads to a commensurate density wave while preserving phase rigidity, thereby enabling supersolidity at the boundary.

The absorption and reflection of superfluid excitations at a solid boundary have long been studied in the context of superfluid helium transport~\cite{Gross1963, Adamenko2008, Pomeau2008, Roberts2009, Ambrosetti2022, Ambrosetti2023}.
These processes can be recast as mode couplings between the superfluid's and solid's fluctuations, as previously established for solid-solid~\cite{Pendry1997, Persson1998, Volokitin1999, Volokitin2006} and solid-liquid interfaces~\cite{Kavokine2022, Coquinot2023b, Bui2023, Lizee2024}.
High-frequency mode coupling usually dominates attractive and repulsive forces like the van der Waals interactions. 
In contrast, the low-frequency quasiparticles like phonons and rotons, which are energetically more accessible, control the friction.
Such coupling enables a direct exchange of excitations across the interface and gives rise to coupled transport phenomena, such as Coulomb drag effects, which appear to be particularly promising for water-solid interfaces~\cite{Coquinot2023, Coquinot2024, Coquinot2025}.
In the following, we will show that these low-frequency mode couplings not only impact the transport properties but also the ground state of the system.

\begin{figure}
	\centering
	\includegraphics{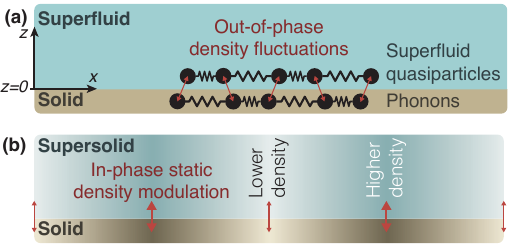}
	\caption{ \tb{Model.}
	\tb{(a)} Schematics of the system: a superfluid is deposited at a smooth solid interface. Both the superfluid and the solid have density fluctuations and they interact through an attractive coupling, but they remain uncorrelated. 
	\tb{(b)} Schematics of a supersolid phase: in order to maximize the attractive interfacial coupling the density of both the superfluid and the solid present an in-phase density modulation in space, breaking the translation symmetry. 	}\label{fig1}
\end{figure}

In this article, we consider interfacial mode coupling between low-frequency density modulations and show that it can drive a supersolid phase transition in both superfluid helium and Bose-Einstein condensates.
The system, sketched in Fig.~\ref{fig1}(a), consists of a (possibly confined) superfluid layer in contact with a perfectly smooth solid, modeled as a continuous elastic medium supporting phonons.
At the interface, the superfluid atoms typically experience van der Waals attraction and Pauli repulsion, which define the characteristic interatomic distance $\sigma$ (typically a few Angstr\"oms) and interaction energy $\mc{E}_0$ (typically of the order of meV) . 
\rev{For simplicity, we model the interfacial interaction by the contact potential
\beq
V_{\rm int}(\r)=-\alpha\delta(\r),
\eeq
although the formalism can also accommodate finite-range or nonlocal interactions.}
Here, we introduced the interfacial coupling constant $\alpha \sim \mc{E}_0\sigma^3$, whose precise definition depends on the form of the solid-superfluid interaction potential.
For realistic systems, we thus expect $\alpha \sim 10^{-2}-10^{-1}$ meV$\cdot$nm$^{3}$.
Note that the coordinate origin was chosen so that the interface $z = 0$ coincides with the positions of both the first superfluid atoms and the solid surface.
{For uniform densities, this interaction shifts the local chemical potential at the mean-field level and may break superfluidity for strong couplings~\cite{Boninsegni2011}.
Here, we investigate whether the response of the solid may lead to density modulations with reasonable couplings.
We therefore focus on density fluctuations and neglect the uniform density shift.}
Thus, we show that above a critical interfacial coupling strength depending on the phononic properties, the coupled densities become unstable toward an oscillatory modulation that maximizes the superfluid-solid overlap, as illustrated in Fig.~\ref{fig1}(b).

\begin{figure*}
	\centering
	\includegraphics[width=\textwidth]{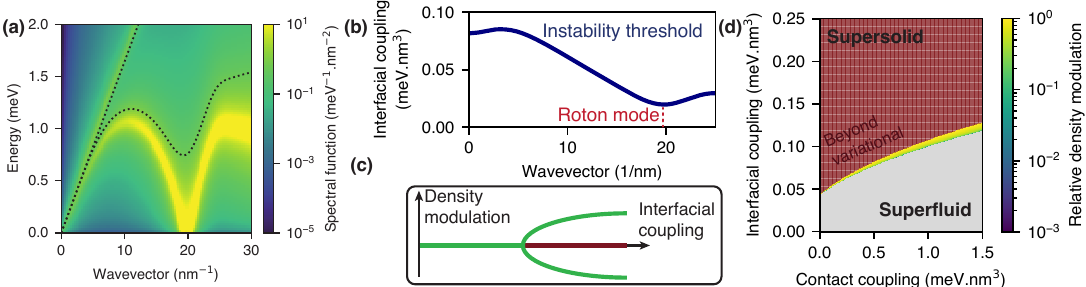}
	\caption{ \tb{Supersolid transition in 2D.}
	\tb{(a)} Spectral function $-\frac{1}{\pi}\im{G^{\rm R}(\q,\w+i0^+)}$ of a 2D layer of He II renormalised by acoustic phonon of speed $c=250$ m/s through an interfacial coupling $\alpha= 0.02$ meV.nm$^3$. 
	Helium has a density of $n_0=2.2\e{28}$ /m$^3$ and we use $\theta=0.3$ nm. {As an example, we use the properties of graphene in the following: an atomic density $ n_{\rm sol}^0=38$ /nm$^2$ and mass $m_{\rm sol}=12\;m_0$  ($m_0$ being the atomic mass).  }
	The dashed lines correspond to the energy bands of non-interacting superfluid helium and phonons. 
	\tb{(b)} Critical interfacial coupling $\alpha$ to induce the supersolid instability as a function of the wavevector of the density modulation for the same system. The minimum corresponds to the roton mode.
	\tb{(c)} Sketch of the instability bifurcation: the uniform density state becomes instable above the interfacial coupling critical and a density modulation then becomes stable. The translation symmetry breaking results into a bifurcation.
	\tb{(d)} Phase diagram for a 2D layer of a BEC of thickness $\theta=1$ nm over a length $L_x=50$ nm interacting with acoustic phonons of speed $c=50$ m/s: relative density modulation amplitude $\epsilon_0/\sqrt{n_0}$ as a function of the contact coupling $g$ and interfacial coupling $\alpha$. The supersolid phase corresponds to a nonzero modulation. The BEC molecular mass $m=7\; m_0$ with density $n_0=10^{23}$ /m$^3$.
	}\label{fig2}
\end{figure*}

We model the superfluid by a Gross--Pitaevskii action:
\beqa 
\mc{S}_{\rm GP} &=&  \int \d t\,\d\r \,  \psi^*(\r,t) \left(i\hbar\partial_t +\frac{\hbar^2}{2m}\Delta+\mu\right)\psi(\r,t) \nonumber\\
&&\hspace{-0.2cm} -\frac{1}{2}\int\d t\,\d\r\,\d\r'\, |\psi(\r',t)|^2V_0(|\r-\r'|) |\psi(\r,t)|^2 \label{S_GP}
\eeqa
where $\psi$ is the superfluid order parameter, $\mu$ the chemical potential and $V_0$ the inter-atomic interaction. 
The minimization of the action yields the bulk density $n_0=\mu/\tilde V_0(\q=0)$ and we define accordingly the healing length $\xi=\hbar/\sqrt{2m\mu}$. 
Linearizing around the uniform solution with $\psi(\r,t)=\sqrt{n_0+\delta n(\r,t)}$ yields the Bogoliubov dispersion for the low-energy density modes,
\beq \mc{E}_{\rm B}(\q)=\sqrt{\mc{E}_{\rm k}(\q)\left(\mc{E}_{\rm k}(\q)+ 2\tilde V(\q)n_0\right) }\eeq
where $\mc{E}_{\rm k}(\q)= \hbar^2\q^2/2m$ is the kinetic energy. 
 In the following, we consider either a BEC with contact interaction $V_0(\r)=g\delta(\r)$ or superfluid helium where $V_0$ is adjusted to reproduce the experimental dispersion.
{On the other hand, the solid density fluctuations $\delta n_{\rm sol}$ are modelled by a Gaussian Keldysh action implementing the susceptibility
$\chi^{\rm R}_{\rm sol}(\q,\w)=[(qc)^2/K_{\rm sol}]/[\w^2-2i\gamma\w-(qc)^2]$
corresponding to an ideal acoustic phonons with sound velocity $c$ and damping $\gamma$. 
Generalisation to more realistic phonon branches is straightforward.}
The static elastic stiffness $K_{\rm sol}=m_{\rm sol} c^2/ n_{\rm sol}^0$ is obtained from the sound velocity $c$ and the average density $n_{\rm sol}^0$ of atoms of mass $m_{\rm sol}$ on the solid surface.
Details of the model and full derivations are provided in the Supplemental Material.

We start by considering a 2D superfluid layer of thickness $\theta\ll \xi$. 
For small density modulations, the Gross--Pitaevskii action can be linearized into a Gaussian form around the bulk density.
Including the interfacial coupling, we can compute the first order self-energy for the superfluid density fluctuations from the product of the superfluid and solid density susceptibilities. 
Thus, the renormalized superfluid susceptibility reads:
\beq \tilde \chi_{n^{\rm 2D}}^{\rm R}(\q,\w) =  \frac{2\theta n_0\mc{E}_{\rm k}(q)}{(\hbar\w)^2-\mc{E}_{\rm B}(q)^2-2\alpha^2n_0\mc{E}_{\rm k}(q) \chi^{\rm R}_{\rm sol}(\q,\w)/\theta}\eeq 
Then, we deduce the hybridized branches of energy for which the denominator of the renormalized susceptibility vanishes.
Due to the interfacial coupling, the upper branch is energetically more costly while the lower branch is softened. 
When the latter becomes negative the corresponding wavevector $\q$ is dynamically unstable.
This happens for
\beq
\alpha^2>\frac{ K_{\rm sol} K_{\rm GP}(q)}{2n_0^2}
\label{eq:alpha2D}
\eeq
where we introduced the stiffness of the superfluid $K_{\rm GP}(q)$, which reads for a 2D layer of thickness $\theta$:
\beq
 K_{\rm GP}^{\rm 2D}(q)=\theta n_0[\mc{E}_{\rm k}(q)+2n_0\tilde V(q)]
 \label{KGP2D}
\eeq
The superfluid stiffness represents the energetic cost, coming from both the kinetic energy and the interatomic interactions, of deforming the density field. 
We find that the instability is favored if the superfluid and the solid have a lower stiffness, corresponding to a lower energy cost for the density modulation of the supersolid phase.
In particular, supersolidity is facilitated by the presence of soft solids, exhibiting slow acoustic phonons.
For instance for superfluid helium, we find reasonable values for the critical interfacial coupling constant for phonons of speed $c=250$ m/s.
Its spectral function (which highlights the branches of energy) close to the threshold is given in Fig.~\ref{fig2}(a): the roton mode is softened and is the first mode to become unstable. 
Consistently, Fig.~\ref{fig2}(b) shows for every wavevector the critical interfacial coupling required to become unstable where the minimum corresponds to the roton mode.
{While we may expect corrections due to the molecular structure of the solid at long wavevectors, the critical wavevector is in general incommensurate with the solid's lattice.}

This mechanism is similar to the one observed in dipolar gases at the onset of supersolidity. 
Therefore, we can test this hypothesis by evaluating the ground-state energy using the variational ansatz $\psi(\r,t) = \sqrt{n_0}\left(1 +  \epsilon_0\cos(q_0x)\right)$ and $\delta n_{\rm sol}(\r,t) = \epsilon_{\rm sol} \cos(q_0x)$ where $x$ is the in-plane direction and $q_0$ is a fixed wavevector.
{Here, we assumed a 1D density modulation for simplicity, as a lattice structure would not change qualitatively our results~\cite{Zhang2019}.}
We use the amplitude of the density modulations $\epsilon_0$ and $\epsilon_{\rm sol}$, which vanish in the superfluid phase and are nonzero in the supersolid phase, as variational parameters.
By plugging this ansatz into the action for the whole system (superfluid and solid), we obtain the energy functional to minimize over the variational parameters. 
The interfacial coupling energy reads
\beq
\mc{E}_{\rm el}+\mc{E}_{\rm int} =\frac{1}{4}K_{\rm sol}\epsilon_{\rm sol}^2-\alpha n_0\epsilon_0\epsilon_{\rm sol}
\label{E_interface}
\eeq
Minimizing with respect to the solid's density modulation yields $ 
\epsilon_{\rm sol} = 2\alpha n_0K_{\rm sol}^{-1} \epsilon_0$.
Thus, the solid modulation is in-phase with the superfluid's modulation. 
{Note, however, that this modulation is independent of the solid’s lattice.}
Plugging this into Eq.~\eqref{E_interface} and including the Gross--Pitaevskii contribution from Eq.~\eqref{S_GP} provides the effective energy cost for the superfluid which reads, up to fourth order in $\epsilon_0$,
\beq
\mc{E}_{\rm eff}[\epsilon_0]  = \frac{1}{2} \left[K_{\rm GP}^{\rm 2D}(q_0)-2\alpha^2n_0^2 K_{\rm sol}^{-1}\right] \epsilon_0^2 +\frac{1}{4}K_4^{\rm 2D}(q_0) \epsilon_0^4
\eeq
where $K_{\rm GP}^{\rm 2D}(q_0)$ is the 2D stiffness of the superfluid given in Eq.~\eqref{KGP2D} and the second term describes the interfacial coupling. 
The quadratic coefficient becomes negative with the same threshold as in Eq.~\eqref{eq:alpha2D}, signaling the transition towards a supersolid state with a density modulation. 
The quartic term, originating from interatomic interactions, stabilizes the density modulation and yields an analytic expression for its amplitude in the limit of small modulations.
{This is a standard Landau--Ginzburg functional with a pitchfork bifurcation, as sketched in Fig.~\ref{fig2}(c), and leading to a spontaneous symmetry breaking.}
Above a threshold the uniform superfluid state becomes unstable while the supersolid state--with positive or negative $\epsilon_0$ and more generally with any shift of the density modulation--becomes stable. 

For a BEC with contact interactions, the stiffness $K_{\rm GP}^{\rm 2D}(q_0)$ (Eq.~\eqref{KGP2D}) is a growing function of the wavevector: the smallest wavevectors are therefore the easiest to destabilize. 
{In practice in a finite size system, its length $L_x$ selects the possible wavevectors and sets the first unstable wavevector $q_{\rm min}=2\pi/L_x$.
On the other hand, the interfacial coupling $\alpha$ selects the larger unstable wavevector $q_{\rm max}$, which is independent from the system size $L_x$.
\rev{Thus, all wavevectors $q_{\rm min}<q_0<q_{\rm max}$ are linearly unstable and the resulting density profile must be determined beyond linear order.
For a BEC with finite-range or nonlocal interactions, an intrinsic length scale may provide a preferred modulation wavevector, similarly to the roton softening for helium.} 
The critical coupling (Eq.~\eqref{eq:alpha2D}) to excite a wavevector $q_0$ reads:}
\beq \alpha_c = \frac{\hbar}{2}\sqrt{\frac{\theta K_{\rm sol} }{n_0 m  }\left(q_0^2 +  2\xi^{-2}\right)} \eeq
For a BEC, the density $n_0$ is smaller than for superfluid helium, leading to a weaker superfluid-solid interaction and therefore a higher threshold.  
 The resulting phase diagram in Fig.~\ref{fig2}(d), computed with a softer solid ($c=50$ m/s), shows the boundary between superfluid and supersolid phases as a function of the interfacial coupling $\alpha$ which drives the instability, and the interaction coupling $g$ which tends towards density uniformity. 
For wavelengths larger than the healing length, the critical interfacial coupling scales with $\alpha_c \propto \xi^{-1} \propto g^{1/2}$.
Thus, the stronger the interatomic interaction, the harder it is to destabilize the density.

\rev{
In the presence of a phase twist $\mc{Q}$ along the $x$ direction, we take the order-parameter ansatz $\psi(\r,t) = \sqrt{n_0}\left(1 +  \epsilon_0\cos(q_0x)\right)e^{i\mc{Q}x+i\phi(\r)}$, where $\phi(\r)$ is periodic and ensures current conservation, $\nabla\cdot[n(\r)(\mc{Q}+\nabla\phi(\r)]=0$. Minimization of the energy yields the same transition as in the absence of the twist. The quadratic energy cost associated with the twist defines the superfluid phase stiffness, \beq
D^{\rm 2D}=\partial_\mc{Q}^2\left[\mc{E}(\mc{Q})-\mc{E}(0)\right]|_{\mc{Q}=0}=\frac{\hbar^2n_0}{m}\left( 1 - \epsilon_0^2\right)^{3/2}.
\eeq
which is independent of the specific form of the interaction potential $V$. The corresponding superfluid fraction is
\beq f_s^{\rm 2D} = \frac{m D^{\rm 2D}/\hbar^2}{\langle n(\r)\rangle_\r}=\frac{\left( 1 - \epsilon_0^2\right)^{3/2}}{1 + \epsilon_0^2/2}
\eeq
Thus, $f_s^{\rm 2D}$ decreases continuously from unity as the density modulation develops, but remains positive throughout the regime of validity of the variational approach.
}

\begin{figure}
	\centering
	\includegraphics[]{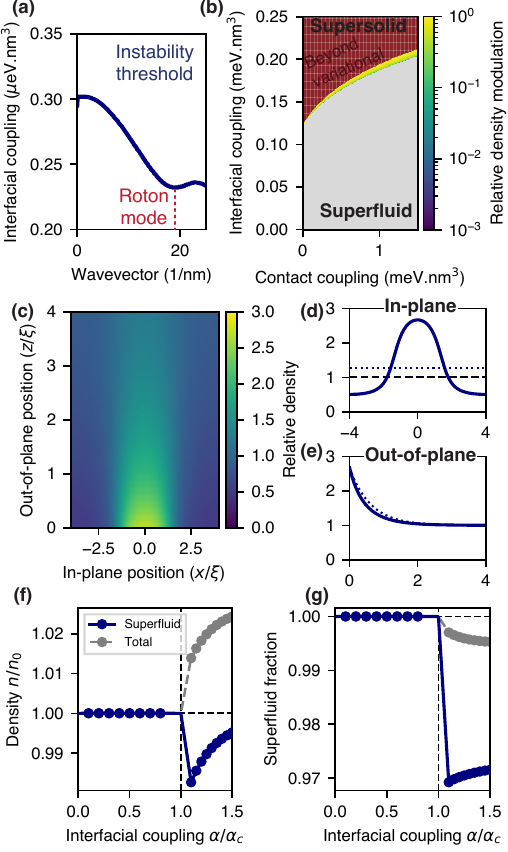}
	\caption{ \tb{Supersolid transition in 3D.}
	\tb{(a)} Critical interfacial coupling $\alpha$ for the supersolid instability versus modulation wavevector for He II coupled to acoustic phonons with $c=250$ m/s. The minimum occurs at the roton mode.
	\tb{(b)} Phase diagram of a BEC of length $L_x=50$ nm coupled to acoustic phonons with $c=50$ m/s: relative modulation amplitude $\epsilon_0/\sqrt{n_0}$ versus contact coupling $g$ and interfacial coupling $\alpha$. The supersolid phase corresponds to a nonzero modulation. 
	\tb{(c)} Ground-state density $n/n_0$ of a BEC for $\alpha=1.1,\alpha_c$, obtained from numerical imaginary-time evolution of the Hamiltonian, with $g=0.75$ meV.nm$^3$,  $\alpha_c\approx 0.16$ meV.nm$^3$,  $\xi\approx 6.3$ nm and $L_x=8\xi=50$ nm.
	\tb{(d)}  Interfacial density profile versus in-plane distance $x/\xi$. Blue and black dashed lines indicate the mean interfacial and bulk densities, respectively.
	\tb{(e)} Density profile normal to the interface at $x=0$ versus $z/\xi$. The dashed line shows the exponential decay predicted in the small-modulation regime.
	\rev{\tb{(f)} Total and superfluid density, extracted from the phase stiffness, versus interfacial coupling.
		\tb{(g)} Corresponding superfluid fraction. The dashed line corresponds to the upper bound $1/\langle n(\r)\rangle_\r \langle \langle n(\r)\rangle_{yz}^{-1}\rangle _x$~\cite{Leggett1970}. }
	}\label{fig3}
\end{figure}

Let us now turn to a 3D system: we consider a half-space superfluid at the solid interface. 
We likewise expect an interfacial instability toward a density modulation which decays far from the interface.
Therefore, we evaluate the ground-state energy using the same variational ansatz where the superfluid density modulation amplitude $\epsilon(z)$ is now a function of the distance to the interface $z$.
Fixing $\epsilon(z=0)=\epsilon_0$, we expect the density modulation amplitude to decay over a skin length $\ell(q_0)$ which replaces the physical thickness $\theta$ of the 2D system. 
Its precise form is obtained by minimizing the Gross--Pitaevskii energy cost which then reduces to:
\beq
\mc{E}_{\rm GP}[\epsilon]=\frac{1}{2}K_{\rm GP}^{\rm 3D}(q_0)\,\epsilon_0^2 + \frac{1}{4}K_{4}^{\rm 3D}(q_0)\,\epsilon_0^4 
\eeq
where $K_{\rm GP}^{\rm 3D}$ is now the 3D stiffness of the superfluid and the stabilizing quartic terms arises from interatomic interactions.

For a BEC with contact interaction, we find that the density modulation amplitude decays exponentially as
$\epsilon(z)=\epsilon_0\,\exp(- z/\ell(q_0))$
where the skin length is $\ell(q_0)=1/\sqrt{q_0^2+2\xi^{-2}}$.
For short wavelengths, the gradients of density are large and therefore the kinetic term dominates the energy cost, which is minimized by a short thickness $\ell(q_0)\approx 1/q_0$.
For long wavelengths however, the interatomic interactions are dominant and the slowly varying density modulation decays over the usual healing length $\ell(q_0)\approx \xi/\sqrt{2}$.
Finally, we obtain the associated 3D stiffness
$ K_{\rm GP}^{\rm 3D}(q_0) = \hbar^2 n_0/2m\ell(q_0)$ which corresponds to the 2D stiffness of a layer of thickness $\ell(q_0)$. 
Since the effective thickness depends on the parameters of the supersolid, the scalings are different.
In particular for long wavelengths the stiffness is less sensitive to the interatomic interaction, $K_{\rm GP}^{\rm 3D}\propto g^{1/2}$, because the thickness of the modulations then shrinks.

For a generic potential $V_0$ (e.g. our model of superfluid helium), we similarly assume an exponential decay $\epsilon(z)=\epsilon_0\,\exp(- z/\ell)$ and determine $\ell(q_0)$ by minimizing the stiffness $K_{\rm GP}^{\rm 3D}(q_0,\ell)$ numerically. 
Including the interfacial contributions, the effective energy cost is again a Landau--Ginzburg functional and the instability threshold is still given by Eq.~\eqref{eq:alpha2D} in which the superfluid stiffness is now 3D.  
Fig.~\ref{fig3}(a) shows the critical interfacial coupling as a function of the wavevector for superfluid helium. 
We again find an instability for reasonable interfacial couplings and first for the roton mode 

For a BEC, the resulting phase diagram in Fig.~\ref{fig3}(b) shows the boundary between superfluid and supersolid phases, with critical coupling
\beq \alpha_c = \frac{\hbar}{2}\sqrt{\frac{K_{\rm sol} }{n_0 m}}\left(q_0^2 + 2 \xi^{-2}\right)^{1/4} \eeq
For wavelengths larger than the healing length, the critical interfacial coupling scales with $ \alpha_c \propto g^{1/4}$, a weaker scaling than in 2D. 
Nevertheless, the condition $\theta \ll \xi$ ensures that the supersolid instability in 3D is less favorable than in 2D, due to the out-of-plane gradient cost. 

\rev{
In the presence of a phase twist, the same variational treatment yields the superfluid phase stiffness and the corresponding superfluid fraction. Unlike in two dimensions, the latter depends on the system size $L_z$ normal to the interface, but remains positive in all cases.
 } 

To prove that the ground state of the system is indeed a supersolid phase beyond the small modulation regime, we compute numerically the ground state of a BEC coupled to the interface by minimizing the total energy.
Integrating out the solid density modulation $\phi$, the interfacial coupling appears as a Robin boundary condition for the bulk Gross--Pitaevskii functional:
\beq
\frac{\hbar^2}{2m}\partial_z \psi|_{z=0} =- \frac{\alpha^2}{ K_{\rm sol}}\,\delta  |\psi|^2\,\psi|_{z=0}
\eeq
where $\delta  |\psi|^2$ is the deviation of the interfacial superfluid density from its average, removing the uniform attraction that is not relevant here.
For strong interfacial couplings, we find a collapse of the superfluid order parameter which focuses at the boundary.
Because this is an interfacial effect, the Lee--Huang--Yang corrections~\cite{Salasnich2018, Skov2021} are not enough to stabilize the superfluid, thus we do not include these higher-order terms. 
However, in practice, the solid cannot accommodate unlimited density accumulation, therefore the right-hand-side term saturates for large superfluid densities.
We thus implement a smooth saturation of the boundary condition in order to avoid any collapse.
At a distance far from the interface $z=z_{\rm max}$, that is far from the interface, the superfluid is homogeneous and thus we impose its density as $\psi(z_{\rm max})=\sqrt{n_0}$. 
{We choose the system size $L_x$ in order to investigate the wavevector $q_0=2\pi/L_x$ and set the range of the interfacial couplings in order to excite only $q_0$.} 

Starting from an initial state that breaks translation symmetry, we then perform an imaginary-time evolution of the Gross--Pitaevskii equation in order to determine the ground state.
As expected, for interfacial couplings below the threshold, the initial density modulation disappears and we find the ground state to be uniform. 
However, above the threshold the density modulation stabilizes with a cosinusoid-like shape, as shown in Fig.~\ref{fig3}(c).
Fig.~\ref{fig3}(d) shows the in-plane interfacial density modulation, which corresponds to the first harmonics of the system, while Fig.~\ref{fig3}(e) displays its out-of-plane decay, which is similar to the exponential decay derived for small modulations. 
\rev{
In the absence of a phase twist, the order parameter remains globally phase coherent. Upon imposing a small phase twist $\mc{Q}$, the phase acquires an additional periodic component that ensures current conservation while preserving global coherence. Numerically, the corresponding increase in the ground-state energy scales quadratically with $\mc{Q}$, allowing us to extract the superfluid phase stiffness and the associated superfluid density and fraction (see Fig.~\ref{fig3}(f--g)). These quantities remain constant below the critical interfacial coupling and decrease once the density-modulated phase emerges, while remaining positive throughout the parameter range considered.}
As anticipated, the ground state of the system therefore corresponds to a superfluid phase below the threshold, and a supersolid phase above it. 

In conclusion, we have shown within a mean-field framework that the elastic response of a smooth interface can promote the transition from a superfluid to a supersolid state. 
\rev{While this coarse-grained description calls for future validation through more advanced numerical approaches like Quantum Monte Carlo (QMC), it isolates a feedback mechanism of the solid that has not been explored in this context~\cite{Wessel2005, Boninsegni2006, Corboz2008, Pollet2008, Boninsegni2011, Bland2022}.}
{Our results therefore suggest an alternative route to realizing supersolids experimentally, based on engineering internal wall properties rather than relying on surface roughness or strong interfacial attraction.
Realizing this scenario experimentally requires a soft solid, supporting slow phonons, and a reasonable interfacial coupling between the solid and the superfluid.}
{Both conditions might be attainable, for instance by engineering metamaterials~\cite{Bertoldi2017, Berger2017} and controlling the solid--superfluid attraction~\cite{Chin2010, Zerba2025}.}
{While we have assumed $T=0$ throughout this paper, incorporating finite-temperature phonon effects would offer a natural and experimentally relevant extension of this mechanism.}
Moreover, our model predicts supersolidity as an interfacial phenomenon while the bulk remains superfluid.
{For helium films, this mechanism may contribute to the phase of the first layers~\cite{Crowell1993, Boninsegni2011, Nyeki2017, Nyeki2017a, Knapp2025}.}
This mixed state could combine the advantages of both phases and their mutual coupling, reminiscent of the cross-term effects between interfacial and bulk transport observed, for example, in classical nanofluidics~\cite{Marbach2019}.
{It could also serve as a seed for bulk supersolid formation by lowering the barrier of the phase transition thanks to an interfacial nucleation.}


\section*{Acknowledgements} 
 B.C. acknowledges support from the NOMIS Foundation. R.\ A. acknowledges funding from the Austrian Academy of Science \"{O}AW grant No. PR1029OEAW03.

\bibliography{bibfile}

\end{document}


\title{\textsc{Supplemental Material} \\Fluctuation-Induced Supersolidity at the Superfluid-Solid Interface }

\author{Baptiste Coquinot, Ragheed Alhyder, Alberto Cappellaro, Mikhail Lemeshko}

\date{\today}

\maketitle


\section{Model}

\subsection{Mean-field description of the superfluid}

At zero-temperature, the order parameter $\psi(\r,t)$ minimizes the Gross-Pitaevskii action:
\beqa 
\mc{S}_{\rm GP} &=&  \int \d t\d\r \,  \psi^*(\r,t) \left(i\hbar\partial_t +\frac{\hbar^2}{2m}\Delta+\mu\right)\psi(\r,t) \nonumber\\
&&\hspace{-0.2cm} -\frac{1}{2}\int\d t\d\r\d\r'\, |\psi(\r',t)|^2V_0(|\r-\r'|) |\psi(\r,t)|^2 \label{S_GP}
\eeqa
where $V_0(\r)$ is the (possibly nonlocal) interatomic interaction. 
$\mu$ is the chemical potential which fixes the bulk superfluid density $n_0$ through
 \beq \mu = n_0\tilde V_0(\q=0) \eeq
 where $\tilde V_0(\q)$ is the Fourier-transform of $V_0(\r)$.
 For a contact interaction $V_0(\r)=g\delta(\r)$ and  $\tilde V_0(\q)=g$. 
 
 To consider fluctuations around $n_0$ we write the order parameter as:
 \beq \psi(\r,t)=\sqrt{n_0+\delta n(\r,t)}e^{i\varphi(\r,t)-i\mu t/\hbar}\approx \sqrt{n_0}\left(1+\frac{\delta n(\r,t)}{2n_0}+i\varphi(\r,t)\right)e^{-i\mu t/\hbar}\eeq
 where $\delta n$ is the density fluctuation and $\varphi$ is the phase fluctuation. 
 For small fluctuations, the Gross-Pitaevskii action reduces to a Gaussian action:
 \beqa 
\mc{S}_{\rm GP}^0[\delta n,\varphi]  &=&- \int \d t\d\r \, \left[\hbar\delta n(\r,t) \partial_t\varphi(\r,t)+\frac{\hbar^2}{2m}n_0[\nabla\varphi(\r,t)]^2 + \frac{\hbar^2}{8mn_0}[\nabla\delta n(\r,t)]^2 \right] \\
&&\ldots -\frac{1}{2}\int\d t\d\r\d\r'\, \delta n(\r',t)V(|\r-\r'|) \delta n(\r,t)
\eeqa
Going to Fourier space,
\beq 
\mc{S}_{\rm GP}^0[\delta n,\varphi]  = -\frac{1}{2}\int \frac{\d \w\d\q}{(2\pi)^{d+1}} \, 
 \begin{pmatrix}
 	\varphi(-\q,-\w) \\ \delta n(-\q,-\w)
 \end{pmatrix}^T
 \begin{pmatrix}
 	\frac{\hbar^2q^2}{m}n_0  & -i\hbar\w   \\ 
 	i\hbar\w & \frac{\hbar^2q^2}{4mn_0} + n_0\tilde V(\q) 
 \end{pmatrix}
  \begin{pmatrix}
 	\varphi(\q,\w) \\ \delta n(\q,\w)
 \end{pmatrix}
\eeq
that we can rewrite as
\beq 
\mc{S}_{\rm GP}^0[\delta n,\varphi]  = \frac{1}{2}\int \frac{\d \w\d\q}{(2\pi)^{d+1}} \, 
 \begin{pmatrix}
 	\varphi(-\q,-\w) \\ \delta n(-\q,-\w)
 \end{pmatrix}^T
\left[ \boldsymbol{\chi}^{\rm R}_{\rm GP}(\q,\w) \right]^{-1}
  \begin{pmatrix}
 	\varphi(\q,\w) \\ \delta n(\q,\w)
 \end{pmatrix}
\eeq
where we introduced the Gross-Pitaevskii retarded susceptibility:
\beq 
\boldsymbol{\chi}^{\rm R}_{\rm GP}(\q,\w)=\begin{pmatrix}
	\chi^{\rm R}_{\varphi}(\q,\w) & \chi^{\rm R}_{\varphi n}(\q,\w) \\
	\chi^{\rm R}_{n\varphi}(\q,\w) & \chi^{\rm R}_{n}(\q,\w)
\end{pmatrix} = \frac{1}{(\hbar\w)^2-\mc{E}_{\rm B}(\q)^2}\begin{pmatrix}
 	\mc{E}_{\rm B}(\q)^2 / 2n_0 \mc{E}_{\rm k}(\q)  & i\hbar\w   \\ 
 	-i\hbar\w &  2n_0 \mc{E}_{\rm k}(\q)
 \end{pmatrix}
 \label{Green-function-GPE}
 \eeq
 with $\mc{E}_{\rm k}(\q)= \hbar^2\q^2/2m$ is the kinetic energy and  
 \beq \mc{E}_{\rm B}(\q)=\sqrt{\mc{E}_{\rm k}(\q)\left(\mc{E}_{\rm k}(\q)+ 2\tilde V(\q)n_0\right) } \eeq
  is the Bogoliubov energy band.
  In particular, the density susceptibility reads:
  \beq 
\chi^{\rm R}_{n}(\q,\w)= \frac{2n_0 \mc{E}_{\rm k}(\q)}{(\hbar\w)^2-\mc{E}_{\rm B}(\q)^2}
 \label{Green-function-GPE2}
 \eeq
 
\subsection{Acoustic phonons}

Let $\u(\r)$ be the in-plane displacement field of the 2D solid at $z=0$. The areal density fluctuation is
\beq
\delta n_{\rm sol}(\r)=-\, n_{\rm sol}^0\,\nabla\!\cdot\!\u(\r),
\label{eq:phi-def}
\eeq
with $n_{\rm sol}^0$ the mean areal atomic density. The elastic Lagrangian density for phonons is:
\beq
\mc L_\u=\frac{\rho_{\rm sol}}{2}\,(\partial_t\u)^2-\frac{\rho_{\rm sol} c^2}{2}\,(\nabla\u)^2,
\eeq
where $\rho_{\rm sol}=m_{\rm sol}n_{\rm sol}^0$ is the areal mass density and $c$ the sound speed. In Fourier space and in terms of $\delta n_{\rm sol}$, we obtain the retarded susceptibility:
\beq
\chi_{\rm sol}^{\rm R}(\q,\w)=K_{\rm sol}^{-1}\frac{(qc)^2}{\w^2-2i\gamma\w-(qc)^2},
\label{eq:GphiR}
\eeq
where we introduced the elastic stiffness 
\beq
K_{\rm sol}=-\chi_{\rm sol}^{\rm R}(\q,0)^{-1}=\frac{m_{\rm sol} c^2}{n_{\rm sol}^0}.
\label{eq:chi-phi-static}
\eeq
and added a relaxation coefficient $\gamma$. 
Finally, the field $\delta n_{\rm sol}$ is described by a Gaussian action:
\beq \mc{S}_{\rm sol}[\delta n_{\rm sol}] = \frac{1}{2} \int \frac{\d\mathbf{q}\d\w}{(2\pi)^3} \, \delta n_{\rm sol}(-\q,-\w) \chi_{\rm sol}^{\rm R}(\q,\w)^{-1}\delta n_{\rm sol}(\q,\w) \eeq

\subsection{Interfacial interaction}

The interactions between a superfluid and a solid is made of two contributions: the van der Waals attraction and the Pauli repulsion.
Typically, these two effect are combined within the effective Lennard-Jones potential  
\beq V_{\rm LJ}(\r)=-\mc{E}_0\left[\left(\frac{\sigma}{r}\right)^{12}-\left(\frac{\sigma}{r}\right)^{6}\right]\eeq
describing the interatomic interaction. 
Here, $\mc{E}_0$ is the interaction energy and $\sigma$ is the typical distance between the atoms. 
Hence,
\beq \mc{H}_{\rm int}(t)=\int\d\r\d\r'\,  |\psi(\r,t)|^2V_{\rm LJ}(\r-\r') n_{\rm sol}(\r',t)\eeq
where $n_{\rm sol}$ is the 2D atomic density of the solid at $z=0$ and $\r'$ runs only over the in-plane dimensions. 
In practice, this interaction is short range so that only the superfluid atoms at a distance $\sim\sigma$ from a given solid atom feel the interactions. 

Thus, in the following, we simplify the problem by taking a simple contact interaction between the solid and the closest atoms of the superfluid. 
Thus,
\beq 
V_{\rm int}(\r)=-\alpha\delta(\r)
\eeq
and 
\beq \mc{H}_{\rm int}(t)=-\alpha\int\d\r\,  |\psi(\r,z=0,t)|^2n_{\rm sol}(\r',t)\eeq
where the integration is over the interface.
Here, we have introduced the coupling constant $\alpha\sim \mc{E}_0\sigma^3$. 
This interaction has an average, which, in mean field, corresponds to increasing the chemical potential at the interface, and thus the superfluid density.
Nevertheless, this cannot induce a supersolid state and therefore is not of interest for this article. 
However, the density fluctuations of both the superfluid and the solid modify the interaction energy, leading to a mode coupling between the superfluid quasiparticles and the solid's phonons.
This is the coupling we want to describe and we therefore consider the interaction Hamiltonian:
\beq \mc{H}_{\rm int}(t)=-\alpha\int\d\r\,  \delta n(\r,z=0,t) \delta n_{\rm sol}(\r',t)\eeq
corresponding to the action:
\beq 
\mc{S}_{\text{int}}[\delta n, \delta n_{\rm sol}] = -\alpha \int \d t \int \frac{\d\mathbf{q}}{(2\pi)^2} \,  \delta n(-\q,z=0,t)\delta n_{\rm sol}(\q,t)
\eeq

\section{Supersolid instability in a 2D layer of superfluid}

\subsection{Renormalised susceptibility}

We consider a 2D layer of superfluid of thickness $\theta$. 
We assume a small $\theta\ll \xi$ in order to neglect out-of-plane fluctuations. We thus have a 2D Gross-Pitaevskii action of bulk density $n_0^{\rm 2D}=\theta n_0$ and density fluctuations $\delta n^{\rm 2D}=\theta\delta n $. While the Gross-Pitaevskii scales with $\theta$, the interfacial action does not. 
Thus, replacing
$n_0$ by $n_0^{\rm 2D}$ and $\delta n$ by $\delta n^{\rm 2D}$ in the following keep the action invariant with an effective coupling constant $\alpha^{\rm 2D}=\alpha/\theta$. In the following of this subsection, we then use the 2D versions without specific notations.

For small fluctuations we can compute the renormalised susceptibility of the superfluid due to the interfacial interaction. 
At first order in self energy, we can obtain it through a Dyson equation, or equivalently by diagonalising the corresponding Gaussian action:
\beq \mc{S}= \frac{1}{2}\int \frac{\d \w\d\q}{(2\pi)^3} \, 
 \begin{pmatrix}
 	\varphi(-\q,-\w) \\ \delta n(-\q,-\w) \\ \delta n_{\rm sol}(-\q,-\w)
 \end{pmatrix}^T
  \begin{pmatrix}
 		 \left[ \boldsymbol{\chi}^{\rm R}_{\rm GP}(\q,\w) \right]^{-1} & -\boldsymbol{\alpha}  \\ 
 -\boldsymbol{\alpha}^T& \left[\chi_{\rm sol}^{\rm R}(\q,\w)\right]^{-1}
 \end{pmatrix}
  \begin{pmatrix}
 	\varphi(\q,\w) \\ \delta n(\q,\w) \\ 	\delta n_{\rm sol}(\q,\w)
 \end{pmatrix}
\eeq
where
\beq  \boldsymbol{\alpha} = \begin{pmatrix} 0 \\  \alpha\end{pmatrix} \eeq
Hence, the renormalised susceptibility of the superfluid reads:
\beq\tilde{ \boldsymbol{\chi}}^{\rm R}_{\rm GP}(\q,\w)= \boldsymbol{\chi}^{\rm R}_{\rm GP}(\q,\w) +  \boldsymbol{\chi}^{\rm R}_{\rm GP}(\q,\w) (s(\q,\w)^{-1} \boldsymbol{\alpha} \otimes \boldsymbol{\alpha} ) \boldsymbol{\chi}^{\rm R}_{\rm GP}(\q,\w) \eeq
where 
\beq s(\q,\w)= \left[\chi_{\rm sol}^{\rm R}(\q,\w)\right]^{-1}- \boldsymbol{\alpha} ^T \boldsymbol{\chi}^{\rm R}_{\rm GP}(\q,\w) \boldsymbol{\alpha}  =  \left[\chi_{\rm sol}^{\rm R}(\q,\w)\right]^{-1}-\alpha^2 \chi^{\rm R}_{n}(\q,\w) \eeq
is the Schur complement and
\beq s^{-1}(\q,\w) = \frac{\chi_{\rm sol}^{\rm R}(\q,\w)}{1-  \alpha^2 \chi_{\rm sol}^{\rm R}(\q,\w)\chi^{\rm R}_{n}(\q,\w)} \eeq
is the renormalised susceptibility of the solid.
Thus,
\beq \tilde{ \boldsymbol{\chi}}^{\rm R}_{\rm GP}(\q,\w)= 
\boldsymbol{\chi}^{\rm R}_{\rm GP}(\q,\w) +  \frac{ \alpha^2\chi_{\rm sol}^{\rm R}(\q,\w)}{1-  \alpha^2 \chi_{\rm sol}^{\rm R}(\q,\w)\chi^{\rm R}_{n}(\q,\w)}
 \begin{pmatrix}
	\chi^{\rm R}_{n\varphi}\chi^{\rm R}_{\varphi n} & \chi^{\rm R}_{\varphi n}\chi^{\rm R}_{n} \\
	\chi^{\rm R}_{n\varphi}\chi^{\rm R}_{n} & [\chi^{\rm R}_{n}]^2
\end{pmatrix}
 \eeq
 \ie 
 \beq \tilde{ \boldsymbol{\chi}}^{\rm R}_{\rm GP}(\q,\w)= 
 \frac{ \boldsymbol{\chi}^{\rm R}_{\rm GP}(\q,\w)}{1-  \alpha^2 \chi_{\rm sol}^{\rm R}(\q,\w)\chi^{\rm R}_{n}(\q,\w)}
 -
   \frac{ \alpha^2\chi_{\rm sol}^{\rm R}(\q,\w)}{1-  \alpha^2 \chi_{\rm sol}^{\rm R}(\q,\w)\chi^{\rm R}_{n}(\q,\w)}
    \begin{pmatrix}
	\tn{det}\left[ \boldsymbol{\chi}^{\rm R}_{\rm GP}\right] & 0 \\
	0 &0
\end{pmatrix}
 \eeq
 Finally, we have found that the superfluid density susceptibility as well as the phase-density correlations are renormalised by a factor $1-  \alpha^2 \chi_{\rm sol}^{\rm R}(\q,\w)\chi^{\rm R}_{n}(\q,\w)$. The renormalised phase fluctuations however, are more subtile, because the solid does not interact directly for the phase fluctuations. As a consequence, the interfacial coupling will not impact significantly this degree of freedom and the phase will remain macroscopically uniform even for beyond the supersolid transition. 
 Using Eq. \eqref{Green-function-GPE} and reestablishing the 3D densities and coupling constants, the renormalised superfluid density susceptibility reads:
 \beq \chi^{\rm R}_{n^{\rm 2D}}(\q,\w) =  \frac{2n_0\theta\mc{E}_{\rm k}(q)}{(\hbar\w)^2-\mc{E}_{\rm B}(q)^2-2\alpha^2n_0\mc{E}_{\rm k}(q) \chi_{\rm sol}^{\rm R}(\q,\w)/\theta}\eeq

\subsection{Hybridized branches of energy and instability criterium}

The effect of the phonon damping $\gamma$ is to broaden the energy branch. Here, we focus on the center of the branch and therefore fix $\gamma=0$.
The effective energy bands of the superfluid are obtained for
\beq (\hbar\w)^2-\mc{E}_{\rm B}(q)^2-2\frac{\alpha^2n_0}{\theta}\mc{E}_{\rm k}(q) \chi_{\rm sol}^{\rm R}(\q,\w) = 0\eeq
corresponding to a vanishing denominator in the renormalised superfluid density susceptibility. 
Using the model of acoustic phonons, we then have:
\beq (\hbar\w)^2-\mc{E}_{\rm B}(q)^2-2\frac{\alpha^2n_0}{\theta}\mc{E}_{\rm k}(q) K_{\rm sol}^{-1}\frac{(qc)^2}{\w^2-(qc)^2} = 0\eeq
that is
\beq [(\hbar\w)^2-\mc{E}_{\rm B}(q)^2][\w^2-(qc)^2]-2\frac{\alpha^2n_0}{\theta} K_{\rm sol}^{-1} \mc{E}_{\rm k}(q)(qc)^2 = 0\eeq
Therefore, we obtain the energy bands:
\beq (\hbar\w)^2=\frac{\mc{E}_{\rm B}(q)^2 + (\hbar qc)^2}{2} \pm \sqrt{\left(\frac{\mc{E}_{\rm B}(q)^2 - (\hbar qc)^2}{2}\right)^2 + 2\frac{\alpha^2n_0}{\theta} K_{\rm sol}^{-1} \mc{E}_{\rm k}(q)(\hbar qc)^2 } \eeq

We obtain an instability if the lower energy becomes negative. 
Thus, we find that $\q$ is unstable if:
\beq \frac{\mc{E}_{\rm B}(q)^2}{\mc{E}_{\rm k}(q)}<2\frac{\alpha^2n_0}{\theta} K_{\rm sol}^{-1} \eeq
We define the superfluid stiffness as:
\beq
 K_{\rm GP}^{\rm 2D}(q)=\theta n_0 \frac{\mc{E}_{\rm B}(q)^2}{\mc{E}_{\rm k}(q)} = \theta n_0[\mc{E}_{\rm k}(q)+2n_0\tilde V(q)]
 \label{KGP2D}
\eeq
so that the instability criterium becomes:
\beq
\alpha^2>\frac{ K_{\rm sol} K_{\rm GP}^{\rm 2D}(q_0)}{2n_0^2}.
\label{eq:alpha2D}
\eeq

\subsection{Variational method for the instability}

We now look for a spatially modulated ground state of the coupled system, motivated by the instability discussed previously. We use the following mean-field ansatz for a supersolid phase:
\begin{align}
\psi(\r,t) &= \sqrt{n_0}\left(1 +  \epsilon_0\cos(q_0x)\right), \\
\delta n_{\rm sol}(\r,t) &= \epsilon_{\rm sol} \cos(q_0x).
\end{align}
where $\epsilon_0$ and $\epsilon_{\rm sol}$ are treated as variational parameters.

This ground state is static, independent on $y$ and periodic in $x$. We then compute the energy cost per unit area for the system.
For the superfluid density, the energy cost is:
\beqa
\mc{E}_{\rm GP}[\epsilon_0] &=& -\mu \theta n_0\left(1 +\frac{1}{2}\epsilon_0^2 \right)+  \frac{1}{2}\theta n_0\epsilon_0^2\frac{\hbar^2}{2m}q_0^2\\
&&\ldots +\frac{\theta n_0^2}{2L_x}\int\d x\d x'\, \left(1 +  \epsilon_0\cos(q_0x)\right)^2V(x-x', q_y=0) \left(1 +  \epsilon_0\cos(q_0x')\right)^2 \nonumber
\eeqa
and thus: 
\beq
\mc{E}_{\rm GP}[\epsilon_0]  =-\frac{\mu  n_0}{2}\theta + \frac{1}{2}K_{\rm GP}^{\rm 2D}(q)\epsilon_0^2+\frac{1}{4}K_4^{\rm 2D}\epsilon_0^4 
\eeq
where $K_{\rm GP}^{\rm 2D}(q)$ was introduced in Eq. \eqref{KGP2D} and indeed corresponds to the superfluid density stiffness, and
\beq K_4^{\rm 2D} =  \frac{\theta n_0^2}{4}\left(2\tilde V(0)  +\tilde V(2q_0) \right)
\eeq
For the phonons, the elastic energy is:
\beq
\mc{E}_{\rm el}[\epsilon_{\rm sol}]  =\frac{K_{\rm sol}\epsilon_{\rm sol}^2}{4}
\eeq
The phonon damping $\gamma$ plays no role here because the ansatz is static. 
Finally, the interfacial coupling bring a negative contribution:
\beq
\mc{E}_{\rm int}[\epsilon_0, \epsilon_{\rm sol}]  = -\alpha n_0\epsilon_0\epsilon_{\rm sol}
\eeq

We first optimize over $\epsilon_{\rm sol}$ for get:
\beq \epsilon_{\rm sol} = 2\alpha n_0K_{\rm sol}^{-1} \epsilon_0 \eeq
Finally the energy functional for $\epsilon_0$ reads:
\beq \mc{E}_{\rm eff}[\epsilon_0]  =  \frac{1}{2} \left[K_{\rm GP}^{\rm 2D}(q_0)-2\alpha^2n_0^2 K_{\rm sol}^{-1}\right] \epsilon_0^2 +\frac{1}{4}K_4^{\rm 2D}(q_0) \epsilon_0^4
\eeq
We recover the instability threshold:
\beq \alpha^2>\frac{ K_{\rm sol} K_{\rm GP}^{\rm 2D}(q_0)}{2n_0^2} \eeq
and the supersolid state which minimizes the energy is:
\beq \epsilon_0 =\sqrt{\frac{2\alpha^2n_0^2 K_{\rm sol}^{-1}-K_{\rm GP}^{\rm 2D}(q_0)}{K_4^{\rm 2D}(q_0)}}
= 2\sqrt{\frac{ 2\alpha^2 n_0 - \theta K_{\rm sol}[\mc{E}_{\rm k}(q)+2n_0\tilde V(q)]}{ n_0 \theta K_{\rm sol} [\tilde V(0) +\frac{1}{2} \tilde V(2q_0)]}} \eeq

For a BEC with contact interaction, we then have
\beq \alpha_c^2 =\frac{ K_{\rm sol} \theta }{2n_0}\left[\frac{\hbar^2q_0^2}{2m}+2n_0g\right]=\frac{ \hbar^2 K_{\rm sol} \theta }{4n_0m} \left[q_0^2+2 \xi^{-2}\right]\eeq
and 
\beq \epsilon_0 
= 2\sqrt{\frac{ 2\alpha^2 n_0 - \theta K_{\rm sol}\left[\frac{\hbar^2q_0^2}{2m}+2n_0g\right]}{ \frac{3}{2} n_0 \theta K_{\rm sol}g}} \eeq

\section{Supersolid instability in a 3D superfluid close to the interface}

\subsection{Variational method for the decay from the interface}

We now apply the variational method to a 3D superfluid close to the interface. We use the following mean-field ansatz for a supersolid phase:
\begin{align}
\psi(\r,t) &= \sqrt{n_0}\left(1 +  \epsilon(z)\cos(q_0x)\right), \\
\delta n_{\rm sol}(\r,t) &= \epsilon_{\rm sol} \cos(q_0x).
\end{align}
with $\epsilon_{\rm sol}$ a variational parameter and $ \epsilon(z)$ a variational functionwith $ \epsilon(z=0)=\epsilon_0$. 

We first optimize the form of the function $ \epsilon(z)$. Keeping only quadratic terms, the energy cost of the supersolid wavefunction per unit volume reads:
\beq
\Delta \mc{E}^{(2)}_{\rm GP}[\epsilon]  =\int_{0}^{+\infty}\!\d z\;\frac{\hbar^2 n_0}{4m}\Big[(\partial_z\epsilon)^2+q_0^2\epsilon^2\Big]
\;+\;n_0^2\int_{0}^{+\infty}\!\d z\d z'\;\epsilon(z)\,V(q_0,z-z')\,\epsilon(z')
\label{eq:E2-functional}
\eeq
The resulting Euler-Lagrange equation provides an equation on the form of $\epsilon(z)$:
\beq
-\frac{\hbar^2 n_0}{2m}\,\partial_z^2\epsilon(z)
+\frac{\hbar^2 n_0}{2m}\,q_0^2\,\epsilon(z)
+2n_0^2\int_{0}^{+\infty}\!\d z'\;V(q_0,z-z')\,\epsilon(z')=0.
\label{eq:EL-bulk}
\eeq
Once this equation is solved we then obtain the 3D superfluid surface stiffness $K_{\rm GP}^{\rm 3D}$ through
\beq
\Delta \mc{E}^{(2)}_{\rm GP}[\epsilon]\;=\;\frac{1}{2}K_{\rm GP}^{\rm 3D}(q_0)\,\epsilon_0^2
\eeq

\paragraph{Contact interaction.}
For \(V(\q,z)=g\,\delta(z)\), the equation for $\epsilon(z)$ becomes:
\beq
-\frac{\hbar^2 n_0}{2m}\,\epsilon''(z)
+\left[\frac{\hbar^2 n_0}{2m}\,q_0^2+2n_0^2 g\right]\epsilon(z)=0,
\eeq
that we rewrite as:
\beq
\epsilon''(z)-\left[q_0^2+2 \xi^{-2}\right]\epsilon(z)=0,
\eeq
thus
\beq
\epsilon(z)=\epsilon_0\,e^{-z/\ell(q_0)},\qquad
\ell(q_0)=\frac{1}{\sqrt{q_0^2+2\xi^{-2}}} \eeq
with 
\beq
\xi=\frac{\hbar}{\sqrt{2m g n_0}}.
\eeq
the healing length.
Finally, the quadratic energy cost becomes:
\beq
\mc E_{\rm GP}^{(2)}[\epsilon_0]
=\epsilon_0^2\left[
\frac{\hbar^2 n_0}{8m}\Big(\ell^{-1}+\ell\ q_0^2 \Big)
+\frac{\ell}{2} n_0^2 g
\right]
=
\frac{\hbar^2 n_0}{4m\ell(q_0)} \epsilon_0^2\label{eq:E2-explicit-contact}
\eeq
so that
\beq
K_{\rm GP}^{\rm 3D}(q_0)
=\frac{\hbar^2 n_0}{2m\ell(q_0)}
\label{KGPContact}
\eeq

\paragraph{Generic potential.}
For a generic interaction $V(\r)$ we keep the exponential decay $\epsilon(z)=\epsilon_0 e^{- z/\ell}$ as a variational ansatz. 
The convolution in the action evaluates to
\beq
\int_{0}^{+\infty}\d z\int_{0}^{+\infty}\d z'\;\epsilon(z)\,V(q_0,z-z')\,\epsilon(z')
=\frac{\epsilon_0^2 \ell}{2}\,\bar V(q_0,\kappa=1/\ell),
\eeq
with
\beq
\bar V(q,\kappa)\equiv\int_{-\infty}^{+\infty}\d s\,e^{-\kappa|s|}\,V(q,s)
=\int_{-\infty}^{+\infty}\frac{\d k_z}{2\pi}\;\tilde V(q,k_z)\,\frac{2\kappa}{\kappa^2+k_z^2}.
\label{eq:barV-def}
\eeq
a Lorentzian averaging of the Fourier-transformed interaction potential.
Finally, we obtain:
\beq
K_{\rm GP}^{\rm 3D}(q_0;\ell)
=\frac{\hbar^2 n_0}{4m}\Big(\ell^{-1}+\ell q_0^2\Big)
+\ell n_0^2\,\bar V(q_0,\ell^{-1}).
\label{eq:E2-explicit-general}
\eeq
The optimal skin length $\ell(q_0)$ is obtained by minimizing $K_{\rm GP}^{\rm 3D}(q_0;\ell)$ with respect to $\ell>0$; using this $\ell(q_0)$ defines $K_{\rm GP}^{\rm 3D}(q_0)\equiv K_{\rm GP}(q_0;\ell(q_0))$.

\paragraph{Sanity check for contact interaction.} For contact interaction $\bar V(q_0,\kappa)= \tilde V(q_0,i\kappa)=g$. Then,
\beq
K_{\rm GP}^{\rm 3D}(q_0;\ell)
=\frac{\hbar^2 n_0}{4m}\Big(\ell^{-1}+\ell q_0^2\Big)
+\frac{\ell n_0^2}{2}g = \frac{\hbar^2 n_0}{4m}\left(\ell^{-1}+\ell \left(q_0^2+2\xi^{-2}\right)\right)
\eeq
which is indeed minimal for $\ell(q_0)=1/\sqrt{q_0^2+2\xi^{-2}}$ and recovers Eq. \eqref{KGPContact}.

\subsection{Variational method for the instability}

We now complete the variational analysis by including the interfacial coupling to the solid and by keeping, beyond quadratic order, the stabilizing quartic term generated by the interaction potential. The Gross-Pitaevskii action provides the volume energy cost:
\beq
\mc{E}_{\rm GP}[\epsilon]=\frac{1}{2}K_{\rm GP}^{\rm 3D}(q_0)\,\epsilon_0^2 + \frac{1}{4}K_{4}^{\rm 3D}(q_0)\,\epsilon_0^4 
\label{EGP}
\eeq
where $K_{\rm GP}^{\rm 3D}(q_0)$ was defined in the previous subsection.
The quartic term arises from interatomic interactions inside the superfluid: 
\beq
\mc E_{\rm GP}^{(4)}[\epsilon]
=\frac{n_0^2}{8}\int_0^\infty\!\d z\!\int_0^\infty\!\d z'\;\epsilon(z)^2\Big[V(q=0,z-z')+\tfrac{1}{2}V(q=2q_0,z-z')\Big]\epsilon(z')^2.
\label{eq:E4-functional-ss}
\eeq
With \(\epsilon(z)=\epsilon_0 e^{- z/\ell}\), we the have:
\beq
\mc E_{\rm GP}^{(4)}[\epsilon_0]
=\frac{\ell n_0^2\,\epsilon_0^4}{32}\left[\bar V(q=0,\kappa=2/\ell)+\frac{1}{2}\,\bar V(q=2q_0,\kappa=2/\ell)\right],
\label{eq:E4-closed}
\eeq
where \(\bar V\) was defined in Eq. \eqref{eq:barV-def}. Thus,
\beq K_4^{\rm 3D}(q_0) = \frac{\ell n_0^2\,\epsilon_0^4}{8}\left[\bar V(q=0,\kappa=2/\ell)+\frac{1}{2}\,\bar V(q=2q_0,\kappa=2/\ell)\right] \eeq

As in the previous section, the elastic energy of the solid mode of amplitude \(\epsilon_{\rm sol}\) and its interfacial coupling to the superfluid read
\beq
\mc E_{\rm el}[\epsilon_{\rm sol}]=\frac{K_{\rm sol} \epsilon_{\rm sol}^2}{4},\qquad
\mc E_{\rm int}[\epsilon,\epsilon_{\rm sol}]= -\alpha\,n_0\,\epsilon_0\,\epsilon_{\rm sol}.
\label{eq:Eel-Eint}
\eeq
Eliminating \(\epsilon_{\rm sol}\) variationally, \(\partial_{\epsilon_{\rm sol}}(\mc E_{\rm el}+\mc E_{\rm int})=0\), gives
\beq
\epsilon_{\rm sol}=2\alpha n_0 K_{\rm sol}^{-1}\,\epsilon_0
\quad\Rightarrow\quad
 \mc E_{\rm el}+\mc E_{\rm int}=-\alpha^2 n_0^2 K_{\rm sol}^{-1}\,\epsilon_0^2
\label{eq:surface-penalty}
\eeq

Combining Eqs. \eqref{EGP} and \eqref{eq:E4-closed}, the energy per unit area as a function of \(\epsilon_0\) reads
\beq
\mc{E}_{\rm eff}[\epsilon_0]  = \frac{1}{2} \left[K_{\rm GP}^{\rm 3D}(q_0)-2\alpha^2n_0^2 K_{\rm sol}^{-1}\right] \epsilon_0^2 +\frac{1}{4}K_4^{\rm 3D}(q_0) \epsilon_0^4
\eeq
which has the same form than in 2D. Therefore, we have the same instability criterium:
\beq
\alpha^2>\frac{ K_{\rm sol} K_{\rm GP}^{\rm 3D}(q)}{2n_0^2}
\label{eq:alpha3D}
\eeq
Beyond this threshold, the optimal amplitude is
\beq \epsilon_0 =\sqrt{\frac{2\alpha^2n_0^2 K_{\rm sol}^{-1}-K_{\rm GP}^{\rm 2D}(q_0)}{K_4^{\rm 2D}(q_0)}}
 \eeq

\paragraph{BEC with contact interaction.} In this case, we have explicitly 
\beq
\alpha_c^2 =\frac{ \hbar^2 K_{\rm sol} }{4n_0m\ell(q_0)}= \frac{ \hbar^2 K_{\rm sol} }{4n_0m}\sqrt{q_0^2+2\xi^{-2}}
\label{eq:alpha-th}
\eeq

\section{Superfluid phase stiffness and superfluid fraction}
\label{sec:superfluid_stiffness}

We now determine whether the spatially modulated state obtained above retains a finite superfluid response. We impose a phase twist $\mc{Q}$ in the $x$-direction by writing the order parameter as
\begin{equation}
\widetilde{\psi}_{\mc{Q}}(\mathbf{r})
=
\sqrt{n_{\mc{Q}}(\mathbf{r})}
\exp\left\{
i\left[
\mc{Q}x+\phi_{\mc{Q}}(\mathbf{r})
\right]
\right\}.
\label{eq:twisted_order_parameter}
\end{equation}
This is equivalent to the replacement $\partial_x\rightarrow\partial_x+iQ$ in the Gross--Pitaevskii functional. The function $\phi_\mc{Q}(\mathbf{r})$ is periodic in $x$ and is chosen such that
\begin{equation}
\left\langle
\partial_x\phi_\mc{Q}
\right\rangle_x
=
0,
\label{eq:periodic_phase_condition}
\end{equation}
where $\langle\cdots\rangle_x$ denotes an average over one period of the density modulation. The total phase winding is therefore fixed by $\mc{Q}$. In the uniform phase, $\phi_\mc{Q}=0$. In the modulated phase, a nonzero periodic contribution $\phi_\mc{Q}$ is generally required because a spatially uniform phase gradient does not produce a divergence-free current in an inhomogeneous density profile.

For static configurations, we work with the energy per unit interfacial area. Let $\mc{E}_{0}(\mc{Q})$ denote the variational ground-state energy after all internal degrees of freedom have been minimized at fixed $\mc{Q}$. The superfluid stiffness in the $x$-direction is defined by
\begin{equation}
\mc{E}_{0}(\mc{Q})
=
\mc{E}_{0}(0)
+
\frac{1}{2}D_x\mc{Q}^2
+
O(\mc{Q}^4).
\label{eq:stiffness_definition}
\end{equation}
Because the energy is even under $\mc{Q}\rightarrow-\mc{Q}$, no term linear in $\mc{Q}$ is present. The corresponding areal superfluid density is
\begin{equation}
n_{s}
=
\frac{m}{\hbar^2}D_x.
\label{eq:areal_superfluid_density}
\end{equation}
For a strictly two-dimensional system, $n_{s,x}$ is the usual two-dimensional superfluid density. For a three-dimensional slab, it is the superfluid density integrated over the thickness of the slab.

To obtain the superfluid fraction at fixed chemical potential, the superfluid density must be normalized by the actual areal particle density of the variational state,
\begin{equation}
\langle n \rangle
=
\frac{1}{A}
\int d\mathbf{r}\,
\left|
\widetilde{\psi}_{\mc{Q}=0}(\mathbf{r})
\right|^2,
\label{eq:areal_particle_density}
\end{equation}
where $A$ is the interfacial area. The superfluid fraction is then
\begin{equation}
f_{s}
=
\frac{n_{s}}{\langle n \rangle}
=
\frac{mD_x}{\hbar^2\langle n \rangle}.
\label{eq:superfluid_fraction_definition}
\end{equation}
This normalization is important here because the variational ansatz is used at fixed chemical potential and does not conserve the total particle number.

For a supersolid, we expect $0<f_s<1$. Indeed, $f_s>0$ because a supersolid remains a superfluid. And the constraint $f_s<1$ comes from Leggett criterium:
\beq f_s\leq \frac{1}{\langle n(\r)\rangle_\r  \langle\langle n(\r)\rangle_{yz}^{-1} \rangle_x}\leq 1 \eeq
Thus, for a supersolid, because of the density modulation, the superfluid fraction must be smaller than 1. 

\subsection{Envelope theorem}

We first explain how the variation of the density profile with $\mc{Q}$ enters the superfluid stiffness. Let $\boldsymbol{\lambda}$ denote all time-reversal-even variational parameters, such as the density-wave amplitude $\epsilon_0$, the solid amplitude $\epsilon_0$, or the penetration length $\ell$. The periodic phase $\phi_\mc{Q}$ is first minimized at fixed $\mc{Q}$ and fixed $\boldsymbol{\lambda}$. We define the reduced energy
\begin{equation}
\mathcal{E}(\mc{Q},\boldsymbol{\lambda})
=
\min_{\phi}
E(\mc{Q},\boldsymbol{\lambda},\phi).
\label{eq:reduced_twisted_energy}
\end{equation}
The variational ground-state energy is subsequently obtained by minimizing over the remaining parameters,
\begin{equation}
\mc{E}_{0}(\mc{Q})
=
\mathcal{E}
\left(
\mc{Q},\boldsymbol{\lambda}_{*}(\mc{Q})
\right),
\label{eq:optimized_energy}
\end{equation}
where
\begin{equation}
\left.
\frac{\partial\mathcal{E}}
{\partial\lambda_a}
\right|_{\boldsymbol{\lambda}=\boldsymbol{\lambda}_{*}(\mc{Q})}
=
0.
\label{eq:variational_stationarity}
\end{equation}

Differentiating Eq.~\eqref{eq:optimized_energy} with respect to $\mc{Q}$ gives
\begin{equation}
\frac{d\mc{E}_{0}}{d\mc{Q}}
=
\frac{\partial\mathcal{E}}{\partial \mc{Q}}
+
\frac{\partial\mathcal{E}}{\partial\lambda_a}
\frac{d\lambda_{*,a}}{d\mc{Q}}.
\label{eq:first_total_derivative}
\end{equation}
The second term vanishes because of the stationarity condition in Eq.~\eqref{eq:variational_stationarity}. Therefore,
\begin{equation}
\frac{d\mc{E}_{0}}{d\mc{Q}}
=
\left.
\frac{\partial\mathcal{E}}{\partial \mc{Q}}
\right|_{\boldsymbol{\lambda}=\boldsymbol{\lambda}_{*}(\mc{Q})}.
\label{eq:envelope_first_derivative}
\end{equation}

For the second derivative, the implicit variation of $\boldsymbol{\lambda}_{*}(\mc{Q})$ can contribute in general. Differentiating the stationarity condition with respect to $\mc{Q}$ gives
\begin{equation}
H_{ab}
\frac{d\lambda_{*,b}}{d\mc{Q}}
+
\frac{\partial^2\mathcal{E}}
{\partial\lambda_a\partial \mc{Q}}
=
0,
\label{eq:derivative_stationarity}
\end{equation}
where
\begin{equation}
H_{ab}
=
\frac{\partial^2\mathcal{E}}
{\partial\lambda_a\partial\lambda_b}
\label{eq:variational_hessian}
\end{equation}
is the Hessian evaluated at the variational minimum. If the minimum is locally stable and nondegenerate, $H$ is positive definite and invertible. Equation~\eqref{eq:derivative_stationarity} then gives
\begin{equation}
\frac{d\lambda_{*,a}}{d\mc{Q}}
=
-
\left(H^{-1}\right)_{ab}
\frac{\partial^2\mathcal{E}}
{\partial\lambda_b\partial \mc{Q}}.
\label{eq:variational_response}
\end{equation}
Using this result in the second derivative of Eq.~\eqref{eq:optimized_energy}, we obtain
\begin{equation}
\frac{d^2\mc{E}_{0}}{d\mc{Q}^2}
=
\frac{\partial^2\mathcal{E}}{\partial \mc{Q}^2}
-
\frac{\partial^2\mathcal{E}}
{\partial \mc{Q}\partial\lambda_a}
\left(H^{-1}\right)_{ab}
\frac{\partial^2\mathcal{E}}
{\partial\lambda_b\partial \mc{Q}}.
\label{eq:envelope_second_derivative}
\end{equation}

After the periodic phase has been minimized, all remaining variational parameters are even under time reversal. Consequently, the reduced energy satisfies
\begin{equation}
\mathcal{E}(\mc{Q},\boldsymbol{\lambda})
=
\mathcal{E}(-\mc{Q},\boldsymbol{\lambda})
\label{eq:reduced_energy_even}
\end{equation}
at fixed $\boldsymbol{\lambda}$. Assuming that $\mathcal{E}$ is analytic around $Q=0$, Eq.~\eqref{eq:reduced_energy_even} implies
\begin{equation}
\left.
\frac{\partial^2\mathcal{E}}
{\partial \mc{Q}\partial\lambda_a}
\right|_{Q=0}
=
0.
\label{eq:mixed_derivative_zero}
\end{equation}
Equation~\eqref{eq:envelope_second_derivative} therefore reduces to
\begin{equation}
D_x
=
\left.
\frac{\partial^2\mathcal{E}}
{\partial \mc{Q}^2}
\right|_{
Q=0,\,
\boldsymbol{\lambda}
=
\boldsymbol{\lambda}_{*}(0)
}.
\label{eq:envelope_stiffness}
\end{equation}
Equivalently, the time-reversal-even parameters satisfy
\begin{equation}
\boldsymbol{\lambda}_{*}(\mc{Q})
=
\boldsymbol{\lambda}_{*}(0)
+
O(\mc{Q}^2),
\label{eq:even_variational_parameters}
\end{equation}
so their $\mc{Q}$-dependence modifies the energy only at order $\mc{Q}^4$.

It is essential that the minimization over $\phi_\mc{Q}$ be performed before applying Eq.~\eqref{eq:envelope_stiffness}. Before this minimization, the phase-dependent energy contains the term
\begin{equation}
\frac{\hbar^2}{m}
Q
\int d\mathbf{r}\,
n(\mathbf{r})
\partial_x\phi_\mc{Q}(\mathbf{r}),
\label{eq:phase_cross_term}
\end{equation}
which is linear in both $\mc{Q}$ and $\phi_\mc{Q}$. The full functional is invariant under the combined transformation
\begin{equation}
\mc{Q}\rightarrow-\mc{Q},
\qquad
\phi_\mc{Q}\rightarrow-\phi_\mc{Q},
\label{eq:phase_time_reversal}
\end{equation}
but it is not even in $\mc{Q}$ if $\phi_\mc{Q}$ is held fixed. The phase therefore responds linearly to the twist,
\begin{equation}
\phi_\mc{Q}(\mathbf{r})
=
\mc{Q}\phi_1(\mathbf{r})
+
O(\mc{Q}^3),
\label{eq:odd_phase_response}
\end{equation}
and this response contributes directly to the coefficient of $\mc{Q}^2$. In contrast, the density-wave amplitude, the penetration length, and the solid amplitude respond only at order $\mc{Q}^2$ and do not provide an additional contribution to the stiffness.

\subsection{Two-dimensional case}

As in Sec.~2, all two-dimensional densities and interaction constants are used without an additional superscript. In particular, $n_0$ denotes the uniform areal density of the two-dimensional layer. At fixed $q_0$, we consider the twisted variational ansatz
\begin{equation}
\widetilde{\psi}_\mc{Q}(x)
=
\sqrt{n_0}
\left[
1+\epsilon_0\cos(q_0x)
\right]
\exp\left\{
i\left[
\mc{Q}x+\phi_\mc{Q}(x)
\right]
\right\},
\label{eq:2d_twisted_ansatz}
\end{equation} 
together with
\begin{equation}
\delta n_{\mathrm{sol}}(x)
=
\epsilon_0
\cos(q_0x).
\label{eq:2d_solid_ansatz_twist}
\end{equation}
The superfluid density profile is
\begin{equation}
n(x)
=
n_0
\left[
1+\epsilon_0\cos(q_0x)
\right]^2.
\label{eq:2d_density_profile}
\end{equation}

The phase-dependent part of the energy per unit area is
\begin{equation}
\mc{E}_{\phi}^{2D}
=
\frac{\hbar^2}{2m}
\left\langle
n(x)
\left[
\mc{Q}+\partial_x\phi_\mc{Q}(x)
\right]^2
\right\rangle_x.
\label{eq:2d_phase_energy}
\end{equation}
Variation with respect to $\phi_\mc{Q}$ gives
\begin{equation}
\partial_x
\left\{
n(x)
\left[
\mc{Q}+\partial_x\phi_\mc{Q}(x)
\right]
\right\}
=
0.
\label{eq:2d_current_conservation}
\end{equation}
This is the stationary current-conservation equation. It implies
\begin{equation}
n(x)
\left[
\mc{Q}+\partial_x\phi_\mc{Q}(x)
\right]
=
C,
\label{eq:2d_constant_current}
\end{equation}
where $C$ is independent of $x$. Averaging the phase gradient over one period and using Eq.~\eqref{eq:periodic_phase_condition}, we find
\begin{equation}
Q
=
C
\left\langle
\frac{1}{n(x)}
\right\rangle_x,
\label{eq:2d_twist_constraint}
\end{equation}
and therefore
\begin{equation}
\mc{Q}+\partial_x\phi_\mc{Q}(x)
=
\frac{Q}
{
n(x)
\left\langle n^{-1}\right\rangle_x
}.
\label{eq:2d_relaxed_phase_gradient_general}
\end{equation}

For the density in Eq.~\eqref{eq:2d_density_profile}, the required average is
\begin{equation}
\left\langle
\frac{1}{n(x)}
\right\rangle_x
=
\frac{1}{n_0}
\frac{1}{2\pi}
\int_0^{2\pi}
\frac{du}{\left(1+\epsilon_0\cos u\right)^2}.
\label{eq:2d_inverse_density_integral}
\end{equation}
The standard integral
\begin{equation}
\frac{1}{2\pi}
\int_0^{2\pi}
\frac{du}{1+\epsilon_0\cos u}
=
\frac{1}{\sqrt{1-\epsilon_0^2}}
\label{eq:standard_cosine_integral}
\end{equation}
implies
\begin{equation}
\frac{1}{2\pi}
\int_0^{2\pi}
\frac{du}{\left(1+\epsilon_0\cos u\right)^2}
=
\frac{1}{\left(1-\epsilon_0^2\right)^{3/2}}.
\label{eq:inverse_density_average}
\end{equation}
The relaxed phase gradient is consequently
\begin{equation}
\mc{Q}+\partial_x\phi_\mc{Q}(x)
=
Q
\frac{
\left(1-\epsilon_0^2\right)^{3/2}
}
{
\left[
1+\epsilon_0\cos(q_0x)
\right]^2
}.
\label{eq:2d_exact_phase_gradient}
\end{equation}
Choosing $\phi_\mc{Q}(0)=0$, the periodic phase correction can be written exactly as
\begin{equation}
\phi_\mc{Q}(x)
=
Q
\int_0^x
dx'
\left\{
\frac{
\left(1-\epsilon_0^2\right)^{3/2}
}
{
\left[
1+\epsilon_0\cos(q_0x')
\right]^2
}
-
1
\right\}.
\label{eq:2d_exact_phase}
\end{equation}
The integral over one period vanishes as a consequence of Eq.~\eqref{eq:inverse_density_average}, so Eq.~\eqref{eq:2d_exact_phase} is periodic. Its expansion in the Landau regime is
\begin{equation}
\phi_\mc{Q}(x)
=
-
\frac{2\mc{Q}\epsilon_0}{q_0}
\sin(q_0x)
+
\frac{3\mc{Q}\epsilon_0^2}{4q_0}
\sin(2q_0x)
+
O(\mc{Q}\epsilon_0^3).
\label{eq:2d_phase_expansion}
\end{equation}

Substitution of Eq.~\eqref{eq:2d_relaxed_phase_gradient_general} into Eq.~\eqref{eq:2d_phase_energy} gives the general one-dimensional harmonic-mean result
\begin{equation}
\mc{E}_{\phi}^{2D}
=
\frac{\hbar^2\mc{Q}^2}{2m}
\left\langle
\frac{1}{n(x)}
\right\rangle_x^{-1}.
\label{eq:2d_harmonic_mean_energy}
\end{equation}
For the present variational profile, this becomes
\begin{equation}
\mc{E}_{\phi}^{2D}
=
\frac{\hbar^2n_0}{2m}
\mc{Q}^2
\left(
1-\epsilon_0^2
\right)^{3/2}.
\label{eq:2d_relaxed_phase_energy}
\end{equation}

We next include the density and solid contributions. For a generic density--density interaction potential, the static Gross--Pitaevskii energy per unit area is
\begin{equation}
\mc{E}_{\mathrm{GP}}^{2D}
=
\left\langle
\frac{\hbar^2}{2m}
\left|
\partial_x\widetilde{\psi}_\mc{Q}
\right|^2
-
\mu
\left|
\widetilde{\psi}_\mc{Q}
\right|^2
\right\rangle_x
+
\frac{1}{2L_x}
\int dx\,dx'\,
n(x)
V(x-x',q_y=0)
n(x').
\label{eq:2d_generic_energy_functional}
\end{equation}
The density profile can be decomposed as
\begin{equation}
n(x)
=
n_0
\left(
1+\frac{\epsilon_0^2}{2}
\right)
+
2n_0\epsilon_0\cos(q_0x)
+
\frac{n_0\epsilon_0^2}{2}
\cos(2q_0x).
\label{eq:2d_density_harmonics}
\end{equation}
Using $\mu=n_0\widetilde{V}(0)$, the density-dependent part of Eq.~\eqref{eq:2d_generic_energy_functional} gives, up to fourth order in $\epsilon_0$,
\begin{equation}
\mc{E}_{\mathrm{GP}}^{2D}
=
-
\frac{\mu n_0}{2}
+
\mc{E}_{\phi}^{2D}
+
\frac{1}{2}
K_{\mathrm{GP}}^{2D}(q_0)
\epsilon_0^2
+
\frac{1}{4}
K_4^{2D}(q_0)
\epsilon_0^4,
\label{eq:2d_gp_energy_twist}
\end{equation}
where
\begin{equation}
K_{\mathrm{GP}}^{2D}(q_0)
=
n_0
\left[
E_k(q_0)
+
2n_0\widetilde{V}(q_0)
\right]
\label{eq:2d_quadratic_coefficient_repeated}
\end{equation}
and
\begin{equation}
K_4^{2D}(q_0)
=
\frac{n_0^2}{4}
\left[
2\widetilde{V}(0)
+
\widetilde{V}(2q_0)
\right].
\label{eq:2d_quartic_coefficient_repeated}
\end{equation}
The elastic and interfacial contributions remain
\begin{equation}
\mc E_{\mathrm{el}}
=
\frac{K_{\mathrm{sol}}}{4}
\epsilon_{\rm sol}^2,
\qquad
\mc E_{\mathrm{int}}
=
-
\alpha n_0
\epsilon_{\rm sol}\epsilon_0.
\label{eq:2d_solid_energy_twist}
\end{equation}
The full variational energy is therefore
\begin{equation}
\begin{aligned}
\mc E^{2D}
\left(
\epsilon_{\rm sol},\epsilon_0,\mc Q
\right)
={}&
-
\frac{\mu n_0}{2}
+
\frac{\hbar^2n_0}{2m}
\mc{Q}^2
\left(
1-\epsilon_0^2
\right)^{3/2}
\\
&+
\frac{1}{2}
K_{\mathrm{GP}}^{2D}(q_0)
\epsilon_0^2
+
\frac{1}{4}
K_4^{2D}(q_0)
\epsilon_0^4
+
\frac{K_{\mathrm{sol}}}{4}
\epsilon_{\rm sol}^2
-
\alpha n_0
\epsilon_{\rm sol}\epsilon_0.
\end{aligned}
\label{eq:2d_full_twisted_energy}
\end{equation}

Minimizing over the solid amplitude gives
\begin{equation}
\epsilon_{\rm sol}
=
2\alpha n_0K_{\mathrm{sol}}^{-1}
\epsilon_0.
\label{eq:2d_solid_amplitude_twist}
\end{equation}
After eliminating $\epsilon_{\rm sol}$, the reduced energy is
\begin{equation}
\begin{aligned}
\mc E_{\mathrm{eff}}^{2D}
\left(
\epsilon_0,\mc Q
\right)
={}&
-
\frac{\mu n_0}{2}
+
\frac{\hbar^2n_0}{2m}
\mc{Q}^2
\left(
1-\epsilon_0^2
\right)^{3/2}
\\
&+
\frac{1}{2}
\left[
K_{\mathrm{GP}}^{2D}(q_0)
-
2\alpha^2n_0^2K_{\mathrm{sol}}^{-1}
\right]
\epsilon_0^2
+
\frac{1}{4}
K_4^{2D}(q_0)
\epsilon_0^4.
\end{aligned}
\label{eq:2d_reduced_twisted_energy}
\end{equation}

At $\mc Q=0$, the uniform superfluid is stable when
\begin{equation}
K_{\mathrm{GP}}^{2D}(q_0)
-
2\alpha^2n_0^2K_{\mathrm{sol}}^{-1}
>
0.
\label{eq:2d_uniform_condition}
\end{equation}
Assuming $K_4^{2D}(q_0)>0$, a continuous transition occurs at
\begin{equation}
\alpha_c^2
=
\frac{
K_{\mathrm{sol}}
K_{\mathrm{GP}}^{2D}(q_0)
}
{
2n_0^2
}.
\label{eq:2d_critical_coupling_stiffness_section}
\end{equation}
Above the threshold, the zero-twist variational amplitude is
\begin{equation}
\epsilon_0^2
=
\frac{
2\alpha^2n_0^2K_{\mathrm{sol}}^{-1}
-
K_{\mathrm{GP}}^{2D}(q_0)
}
{
K_4^{2D}(q_0)
}.
\label{eq:2d_ground_state_amplitude_stiffness}
\end{equation}

The phase has already been minimized in Eq.~\eqref{eq:2d_reduced_twisted_energy}. Moreover, $\epsilon_0(\mc{Q})$ is even in $\mc{Q}$. Equation~\eqref{eq:envelope_stiffness} therefore gives
\begin{equation}
D^{2D}
=
\frac{\hbar^2n_0}{m}
\left[
1-
\epsilon_0^2
\right]^{3/2}.
\label{eq:2d_superfluid_stiffness}
\end{equation}
In the uniform superfluid phase, $\epsilon_0=0$, and
\begin{equation}
D_{x,\mathrm{SF}}^{2D}
=
\frac{\hbar^2n_0}{m}.
\label{eq:2d_uniform_stiffness}
\end{equation}
In the supersolid phase,
\begin{equation}
D_{x,\mathrm{SS}}^{2D}
=
\frac{\hbar^2n_0}{m}
\left[
1-
\epsilon_0^2
\right]^{3/2}.
\label{eq:2d_supersolid_stiffness}
\end{equation}

At fixed chemical potential, the actual mean particle density is
\begin{equation}
\langle n \rangle
=
n_0
\left[
1+
\frac{
\epsilon_0^2
}{2}
\right].
\label{eq:2d_mean_density}
\end{equation}
The two-dimensional superfluid fraction is consequently
\begin{equation}
f_{s}^{2D}
=
\frac{
\left[
1-
\epsilon_0^2
\right]^{3/2}
}
{
1+
\epsilon_0^2/2
}.
\label{eq:2d_superfluid_fraction}
\end{equation}
In the Landau regime,
\begin{equation}
f_{s}^{2D}
=
1
-
2
\epsilon_0^2
+
\frac{11}{8}
\epsilon_0^4
+
O
\left[
\epsilon_0^6
\right].
\label{eq:2d_superfluid_fraction_expansion}
\end{equation}
For $\left|\epsilon_0\right|<1$, both the stiffness and the superfluid fraction are strictly positive. In particular, $\epsilon_0$ vanishes continuously at $\alpha=\alpha_c$, so the translational-symmetry-breaking phase immediately above the transition retains a finite superfluid stiffness. It is therefore a supersolid within the present mean-field description.

The derivation of Eqs.~\eqref{eq:2d_superfluid_stiffness} and \eqref{eq:2d_superfluid_fraction} does not rely on a contact interaction. A generic density--density interaction modifies $K_{\mathrm{GP}}^{2D}(q_0)$, $K_4^{2D}(q_0)$, the selected wavevector $q_0$, and the ground-state value of $\epsilon_0$. It does not modify the phase equation in Eq.~\eqref{eq:2d_current_conservation}. For a given density profile, the phase stiffness is always determined by the harmonic mean in Eq.~\eqref{eq:2d_harmonic_mean_energy}, provided that the kinetic energy has the standard Galilean-invariant form and the interaction has no explicit dependence on the phase or current.

\subsection{Three-dimensional case}

We now consider the three-dimensional variational profile introduced in Sec.~3. We retain the exponential form
\begin{equation}
\epsilon(z)
=
\epsilon_0e^{-z/\ell}
\label{eq:3d_exponential_profile_stiffness}
\end{equation}
and write the twisted order parameter as
\begin{equation}
\widetilde{\psi}_\mc{Q}(x,z)
=
\sqrt{n_0}
\left[
1+
\epsilon_0e^{-z/\ell}
\cos(q_0x)
\right]
\exp\left\{
i\left[
\mc{Q}x+\phi_\mc{Q}(x,z)
\right]
\right\}.
\label{eq:3d_twisted_ansatz}
\end{equation}
The density is
\begin{equation}
n(x,z)
=
n_0
\left[
1+
\epsilon_0e^{-z/\ell}
\cos(q_0x)
\right]^2.
\label{eq:3d_density_profile_stiffness}
\end{equation}

The phase-dependent energy per unit interfacial area is
\begin{equation}
\mc{E}_{\phi}^{3D}
=
\frac{\hbar^2}{2m}
\int_0^\infty dz\,
\left\langle
n(x,z)
\left\{
\left[
\mc{Q}+\partial_x\phi_\mc{Q}(x,z)
\right]^2
+
\left[
\partial_z\phi_\mc{Q}(x,z)
\right]^2
\right\}
\right\rangle_x.
\label{eq:3d_phase_energy}
\end{equation}
Variation with respect to $\phi_\mc{Q}$ gives the current-conservation equation
\begin{equation}
\partial_x
\left\{
n
\left[
\mc{Q}+\partial_x\phi_\mc{Q}
\right]
\right\}
+
\partial_z
\left(
n\partial_z\phi_\mc{Q}
\right)
=
0.
\label{eq:3d_current_conservation}
\end{equation}
At the solid interface, the absence of a normal superfluid current imposes
\begin{equation}
\left.
\partial_z\phi_\mc{Q}(x,z)
\right|_{z=0}
=
0.
\label{eq:3d_phase_neumann}
\end{equation}
Far from the interface, the flow must approach the uniform bulk flow, so
\begin{equation}
\lim_{z\rightarrow\infty}
\phi_\mc{Q}(x,z)
=
0.
\label{eq:3d_phase_bulk_boundary}
\end{equation}

In the Landau regime, $\phi_\mc{Q}=O(\mc{Q}\epsilon_0)$. Expanding Eq.~\eqref{eq:3d_current_conservation} to first order in $\epsilon_0$ gives
\begin{equation}
\left(
\partial_x^2+\partial_z^2
\right)
\phi_\mc{Q}(x,z)
=
2Qq_0\epsilon_0
e^{-z/\ell}
\sin(q_0x).
\label{eq:3d_linear_phase_equation}
\end{equation}
The source contains only the first harmonic, and we may write
\begin{equation}
\phi_\mc{Q}(x,z)
=
F_\mc{Q}(z)\sin(q_0x).
\label{eq:3d_phase_harmonic}
\end{equation}
The function $F_\mc{Q}$ obeys
\begin{equation}
F_\mc{Q}''(z)
-
q_0^2F_\mc{Q}(z)
=
2\mc Qq_0\epsilon_0e^{-z/\ell}.
\label{eq:3d_phase_ode}
\end{equation}
For $q_0\ell\neq1$, the decaying solution has the form
\begin{equation}
F_\mc{Q}(z)
=
A e^{-z/\ell}
+
B e^{-q_0z}.
\label{eq:3d_phase_general_solution}
\end{equation}
Equation~\eqref{eq:3d_phase_ode} fixes
\begin{equation}
A
=
\frac{
2\mc Qq_0\epsilon_0\ell^2
}
{
1-q_0^2\ell^2
},
\label{eq:3d_phase_particular_coefficient}
\end{equation}
while Eq.~\eqref{eq:3d_phase_neumann} gives
\begin{equation}
B
=
-
\frac{A}{q_0\ell}.
\label{eq:3d_phase_homogeneous_coefficient}
\end{equation}
The periodic phase correction is therefore
\begin{equation}
\phi_\mc{Q}(x,z)
=
\frac{
2\mc Qq_0\epsilon_0\ell^2
}
{
1-q_0^2\ell^2
}
\left[
e^{-z/\ell}
-
\frac{e^{-q_0z}}{q_0\ell}
\right]
\sin(q_0x)
+
O(\mc{Q}\epsilon_0^2).
\label{eq:3d_phase_solution}
\end{equation}
Although Eq.~\eqref{eq:3d_phase_solution} appears singular at $q_0\ell=1$, this is only a singularity of the chosen representation. The resonant solution is finite and is obtained by taking the limit $q_0\ell\rightarrow1$. We exclude this isolated value in the following.

The first-order result in Eq.~\eqref{eq:3d_phase_solution} is sufficient for the Landau calculation of the stiffness. Indeed, the phase source is linear in $\epsilon_0$, and the phase energy is quadratic in the phase gradients. The relaxation contribution generated by Eq.~\eqref{eq:3d_phase_solution} is therefore of order $\mc{Q}^2\epsilon_0^2$. Corrections to the phase profile contribute only to higher orders in the Landau expansion.

To evaluate the phase energy, we expand Eq.~\eqref{eq:3d_phase_energy} to order
$\mc{Q}^2\epsilon_0^2$ in the Landau expansion. This gives
\begin{equation}
\begin{aligned}
\mc{E}_{\phi}^{3D}
={}&
\frac{\hbar^2n_0}{2m}
\int_0^{L_z}dz\,
\Bigg\langle
\mc{Q}^2
+
\mc{Q}^2\epsilon_0^2
e^{-2z/\ell}
\cos^2(q_0x)
\\
&\hspace{2.5cm}
+
4\mc{Q}\epsilon_0
e^{-z/\ell}
\cos(q_0x)
\partial_x\phi_\mc{Q}
+
\left|
\nabla\phi_\mc{Q}
\right|^2
\Bigg\rangle_x
+
O(\mc{Q}^2\epsilon_0^4,\mc{Q}^4),
\end{aligned}
\label{eq:3d_phase_energy_expanded}
\end{equation}
where $L_z$ is a large auxiliary depth satisfying $L_z\gg\ell$ and $q_0L_z\gg1$. Exponentially small corrections associated with the upper boundary are omitted.

We introduce the dimensionless parameter
\begin{equation}
s
=
q_0\ell.
\label{eq:dimensionless_penetration_parameter}
\end{equation}
Direct substitution of Eq.~\eqref{eq:3d_phase_solution} gives
\begin{equation}
\int_0^\infty dz\,
e^{-z/\ell}F_\mc{Q}(z)
=
-
\mc{Q}\epsilon_0\ell^2
\frac{s+2}{(1+s)^2}.
\label{eq:3d_phase_overlap_integral}
\end{equation}
and
\begin{equation}
\int_0^\infty dz\,
\left\langle
\left|
\nabla\phi_\mc{Q}
\right|^2
\right\rangle_x
=
\mc{Q}^2\epsilon_0^2\ell
\frac{s(s+2)}{(1+s)^2}.
\label{eq:3d_phase_gradient_integral}
\end{equation}
The direct contribution associated with the increased mean density is
\begin{equation}
\int_0^\infty dz\,
\left\langle
e^{-2z/\ell}
\cos^2(q_0x)
\right\rangle_x
=
\frac{\ell}{4}.
\label{eq:3d_direct_density_integral}
\end{equation}
Combining Eqs.~\eqref{eq:3d_phase_energy_expanded}, \eqref{eq:3d_phase_overlap_integral}, \eqref{eq:3d_phase_gradient_integral}, and \eqref{eq:3d_direct_density_integral}, we find
\begin{equation}
\mc{E}_{\phi}^{3D}
=
\frac{\hbar^2n_0}{2m}
\mc{Q}^2
\left[
L_z
+
\epsilon_0^2
C_x(q_0,\ell)
\right]
+
O(\mc{Q}^2\epsilon_0^4,\mc{Q}^4),
\label{eq:3d_relaxed_phase_energy}
\end{equation}
where
\begin{equation}
C_x(q_0,\ell)
=
\ell
\left[
\frac{1}{4}
-
\frac{
q_0\ell
\left(
2+q_0\ell
\right)
}
{
\left(
1+q_0\ell
\right)^2
}
\right].
\label{eq:3d_surface_stiffness_correction}
\end{equation}
The first term in Eq.~\eqref{eq:3d_surface_stiffness_correction} is the rigid contribution of the additional particles at fixed chemical potential. The second term is the reduction caused by the relaxation of the current in the $x$-$z$ plane.

For a generic interaction potential, the zero-twist density energy associated with the exponential profile is determined by the coefficients introduced in Sec.~3,
\begin{equation}
K_{\mathrm{GP}}^{3D}(q_0;\ell)
=
\frac{\hbar^2n_0}{4m}
\left(
\ell^{-1}
+
\ell q_0^2
\right)
+
\ell n_0^2
\overline{V}
\left(
q_0,\ell^{-1}
\right)
\label{eq:3d_quadratic_coefficient_repeated}
\end{equation}
and
\begin{equation}
K_4^{3D}(q_0;\ell)
=
\frac{\ell n_0^2}{8}
\left[
\overline{V}
\left(
0,2/\ell
\right)
+
\frac{1}{2}
\overline{V}
\left(
2q_0,2/\ell
\right)
\right].
\label{eq:3d_quartic_coefficient_repeated}
\end{equation}
The full variational energy per unit area is thus
\begin{equation}
\begin{aligned}
\mc E^{3D}
\left(
\epsilon_{\rm sol},
\epsilon_0,
\ell,
Q
\right)
={}&
-
\frac{\mu n_0}{2}L_z
+
\frac{\hbar^2n_0}{2m}
\mc{Q}^2
\left[
L_z
+
\epsilon_0^2
C_x(q_0,\ell)
\right]
\\
&+
\frac{1}{2}
K_{\mathrm{GP}}^{3D}(q_0;\ell)
\epsilon_0^2
+
\frac{1}{4}
K_4^{3D}(q_0;\ell)
\epsilon_0^4
\\
&+
\frac{K_{\mathrm{sol}}}{4}
\epsilon_{\rm sol}^2
-
\alpha n_0
\epsilon_{\rm sol}\epsilon_0
+
O(\mc{Q}^2\epsilon_0^4,\mc{Q}^4,\epsilon_0^6).
\end{aligned}
\label{eq:3d_full_twisted_energy}
\end{equation}

Minimization over the solid amplitude gives
\begin{equation}
\epsilon_{\rm sol}
=
2\alpha n_0K_{\mathrm{sol}}^{-1}
\epsilon_0.
\label{eq:3d_solid_amplitude_twist}
\end{equation}
The reduced energy becomes
\begin{equation}
\begin{aligned}
\mc E_{\mathrm{eff}}^{3D}
\left(
\epsilon_0,\ell,Q
\right)
={}&
-
\frac{\mu n_0}{2}L_z
+
\frac{\hbar^2n_0}{2m}
\mc{Q}^2
\left[
L_z
+
\epsilon_0^2
C_x(q_0,\ell)
\right]
\\
&+
\frac{1}{2}
\left[
K_{\mathrm{GP}}^{3D}(q_0;\ell)
-
2\alpha^2n_0^2K_{\mathrm{sol}}^{-1}
\right]
\epsilon_0^2
+
\frac{1}{4}
K_4^{3D}(q_0;\ell)
\epsilon_0^4
\\
&+
O(\mc{Q}^2\epsilon_0^4,\mc{Q}^4,\epsilon_0^6).
\end{aligned}
\label{eq:3d_reduced_twisted_energy}
\end{equation}

At $\mc Q=0$, the transition occurs when the quadratic coefficient changes sign. For a fixed $\ell$, the threshold is
\begin{equation}
\alpha_c^2
=
\frac{
K_{\mathrm{sol}}
K_{\mathrm{GP}}^{3D}(q_0;\ell)
}
{
2n_0^2
}.
\label{eq:3d_critical_coupling_stiffness_section}
\end{equation}
Close to the transition, $\ell$ is chosen by minimizing $K_{\mathrm{GP}}^{3D}(q_0;\ell)$ with respect to $\ell>0$. Assuming $K_4^{3D}(q_0;\ell)>0$, the zero-twist supersolid amplitude is
\begin{equation}
\epsilon_0^2
=
\frac{
2\alpha^2n_0^2K_{\mathrm{sol}}^{-1}
-
K_{\mathrm{GP}}^{3D}(q_0;\ell)
}
{
K_4^{3D}(q_0;\ell)
}.
\label{eq:3d_ground_state_amplitude_stiffness}
\end{equation}
The penetration length may also acquire a $\mc{Q}$-dependent correction. However, both $\epsilon_0(\mc{Q})$ and $\ell(\mc{Q})$ are even in $\mc{Q}$, and therefore
\begin{equation}
\epsilon_0(\mc{Q})
=
\epsilon_0
+
O(\mc{Q}^2),
\qquad
\ell(\mc{Q})
=
\ell(Q=0)
+
O(\mc{Q}^2).
\label{eq:3d_even_variational_response}
\end{equation}
After the phase has been eliminated, Eq.~\eqref{eq:envelope_stiffness} applies to both variables. Their implicit $\mc{Q}$-dependence does not provide an additional contribution to the quadratic stiffness.

The superfluid stiffness of a slab of depth $L_z$ is therefore
\begin{equation}
D^{3D}(L_z)
=
\frac{\hbar^2n_0}{m}
\left[
L_z
+
\epsilon_0^2
C_x(q_0,\ell)
\right]
+
O
\left[
\epsilon_0^4
\right].
\label{eq:3d_finite_slab_stiffness}
\end{equation}
At fixed chemical potential, its areal particle density is
\begin{equation}
\begin{aligned}
\langle n \rangle
&=
\int_0^{L_z}dz\,
\left\langle
n(x,z)
\right\rangle_x
\\
&=
n_0
\left[
L_z
+
\frac{
\epsilon_0^2
\ell
}{4}
\right]
+
O
\left(
e^{-2L_z/\ell}
\right).
\end{aligned}
\label{eq:3d_areal_particle_density}
\end{equation}
The finite-slab superfluid fraction is consequently
\begin{equation}
f_{s}^{3D}(L_z)
=
\frac{
L_z
+
\epsilon_0^2
\ell
\left[
\frac{1}{4}
-
\frac{
q_0\ell
\left(
2+q_0\ell
\right)
}
{
\left(
1+q_0\ell
\right)^2
}
\right]
}
{
L_z
+
\epsilon_0^2
\ell/4
}
+
O
\left[
\epsilon_0^4
\right].
\label{eq:3d_finite_slab_fraction}
\end{equation}
Equivalently, close to the transition,
\begin{equation}
f_{s}^{3D}(L_z)
=
1
-
\epsilon_0^2
\frac{\ell}{L_z}
\frac{
q_0\ell
\left(
2+q_0\ell
\right)
}
{
\left(
1+q_0\ell
\right)^2
}
+
O
\left[
\epsilon_0^4
\right].
\label{eq:3d_fraction_expansion}
\end{equation}
For every fixed finite $L_z$, this fraction is positive for a sufficiently small supersolid amplitude. 

In conclusion, in two dimensions, the entire fluid must flow through the density minima, and the reduction of the superfluid fraction is of order $\left(\epsilon_0\right)^2$. In three dimensions, the current can additionally redistribute in the $z$-direction, and the interfacial reduction of the total fraction is diluted by the geometrical factor $\ell/L_z$. For every finite $L_z$, however, the stiffness remains strictly positive at sufficiently small $\epsilon_0$. The phase above the instability threshold therefore combines broken translational symmetry near the interface with a finite superfluid stiffness and is a supersolid interfacial state.

\section{Numerical methods for the ground state}

\subsection{Interfacial couplings as a Robin condition}

For numerical purposes it is convenient to integrate out the solid degree of freedom $\delta n_{\rm sol}$.
We focus on static configurations and drop the time dependence.  
Then, the surface energy reads:
\begin{equation}
  \mc{E}_{\rm el}[\delta n_{\rm sol}] +   \mc{E}_{\rm int}[\psi,\delta n_{\rm sol}]=
  \frac{K_{\rm sol}}{2} \int_{z=0} \d\r\,\delta n_{\rm sol}(\r)^2
  - \alpha \int_{z=0} \d\r\,|\psi(\r)|^2\,\delta n_{\rm sol}(\r),
  \label{eq:Eloc_surf}
\end{equation}
To compute the ground state of the system, we can first minimize the surface energy to obtain  $\delta n_{\rm sol}$ as a function of $\delta n$:
\begin{equation}
  \delta n_{\rm sol}(\r) =
  \frac{\alpha}{K_{\rm sol}}\,\delta|\psi(z=0,\r)|^2.
  \label{eq:delta_n_sol_opt}
\end{equation}
where we removed the surface average in $\delta|\psi(z=0,\r)|^2=|\psi(z=0,\r)|^2-\langle|\psi(z=0,\r)|^2\rangle_\r$ to take into account the constraint $\langle   \delta n_{\rm sol}(\r) \rangle_\r=0$ of solid atom number conservation.
Then, the surface energy reads:
\begin{equation}
  \mc{E}_{\rm el}[\delta n_{\rm sol}] +   \mc{E}_{\rm int}[\psi,\delta n_{\rm sol}]=
   -\frac{\alpha^2}{2K_{\rm sol}} \int_{z=0} \d\r\,\left(\delta|\psi(z=0,\r)|^2\right)^2
  \label{eq:Eloc_surf_minimized}
\end{equation}

Adding this interfacial term to the Gross--Pitaevskii functional with contact interactions leads to the total energy of the system:
\beq
\mc{E}_{\rm tot}[\psi] =  \int_{z>0}\d\r \, \left(\frac{\hbar^2}{2m}|\nabla \psi(\r)|^2 -\mu|\psi(\r)|^2 +\frac{1}{2}g|\psi(\r)|^4 \right)
   -\frac{\alpha^2}{2K_{\rm sol}} \int_{z=0} \d\r\,\left(\delta|\psi(z=0,\r)|^2\right)^2
\eeq
Taking into account the boundaries, the integration by parts of the kinetic term gives:
 \beq  \frac{\hbar^2}{2m}\int_{z>0}\d\r \,|\nabla \psi(\r)|^2 = -\frac{\hbar^2}{2m}\int_{z>0}\d\r \,\psi(\r)^*\Delta  \psi(\r)-\frac{\hbar^2}{2m}\int_{z=0}\d\r \,\psi(\r)^*\partial_z  \psi(\r)
 \eeq
 Thus, the minimization over $\psi^*$ of both the volume and the bulk energy provides the usual Gross--Pitaevskii equation:
 \beq \frac{\hbar^2}{2m}\Delta \psi(\r) + \mu \psi(\r) -g | \psi(\r)|^2 \psi(\r)=0 \eeq
 with a Robin boundary condition at $z=0$:
 \beq \frac{\hbar^2}{2m}\partial_z  \psi(\r)|_{z=0}=-\frac{\alpha^2}{K_{\rm sol}}\delta|\psi(z=0,\r)|^2 \psi(\r)|_{z=0}\eeq
 
In the following, we denote
\beq B_0(\r) =\frac{2m \alpha}{\hbar^2}\delta n_{\rm sol}(\r)=\frac{2m}{\hbar^2} \frac{\alpha^2}{K_{\rm sol}}\delta|\psi(z=0,\r)|^2 \label{eqB}\eeq
In practice, the solid density saturates, so that the magnitude of the
boundary field cannot become arbitrarily large. We model this effect
phenomenologically using a smooth hyperbolic-tangent saturation,
\beq
B(\r)
=
B_{\rm avg}
+
B_{\rm max}
\tanh\left[
\frac{B_0(\r)-B_{\rm shift}}{B_{\rm max}}
\right],
\label{B_sat}
\eeq
where $B_0(\r)$ is given by Eq.~\eqref{eqB}. The spatially uniform
quantity $B_{\rm shift}$ is chosen such that
\beq
\left\langle
\tanh\left[
\frac{B_0(\r)-B_{\rm shift}}{B_{\rm max}}
\right]
\right\rangle_{\r}
=0,
\eeq
thereby imposing $\langle B(\r)\rangle_{\r}=B_{\rm avg}$. For
$|B_0-B_{\rm shift}|\ll B_{\rm max}$, the linear response is recovered,
$B(\r)-B_{\rm avg}\simeq B_0(\r)-B_{\rm shift}$, whereas at large
amplitudes the fluctuating part of the field saturates smoothly at
$|B(\r)-B_{\rm avg}|=B_{\rm max}$.
With this cutoff, the associated energy for the interface then reads:
\beq
E_{\rm int}[\rho_s,B]
=
\int d^2r\,
\frac{\hbar^2}{2m} \frac{K_{\rm sol}}{\alpha^2} B_{\rm max}^2
\left(
y(\r)\operatorname{artanh}y(\r)
+
\frac{1}{2}\ln\left[1-y(\r)^2\right]
\right)
\eeq
where 
\beq 
y(\r)=\frac{B(\r)-B_{\rm avg}}{B_{\rm max}}.
\label{E_interface_sat}
\eeq


For numerical computation we use dimensionless variables. The lengths are normalised by the healing length $\xi = \hbar / \sqrt{2 m g n_0}$ and the order parameter by the bulk density $ \sqrt{n_0}=\sqrt{\mu/ g}$ so that $\psi(\r)\rightarrow 1$ in the bulk. The Gross--Pitaevskii equation then becomes:
 \beq \Delta \psi(\r) +  \psi(\r) - | \psi(\r)|^2 \psi(\r)=0 \eeq
 with a Robin boundary condition at $z=0$:
  \beq \partial_z  \psi(\r)|_{z=0}=-B(r)\psi(\r)|_{z=0}\eeq
  where 
\beq B_0(\r) = \frac{\alpha^2}{K_{\rm sol} g\xi}\delta|\psi(z=0,\r)|^2\eeq

\subsection{Imaginary time integration}

\begin{figure}
\centering
\includegraphics[width=\textwidth]{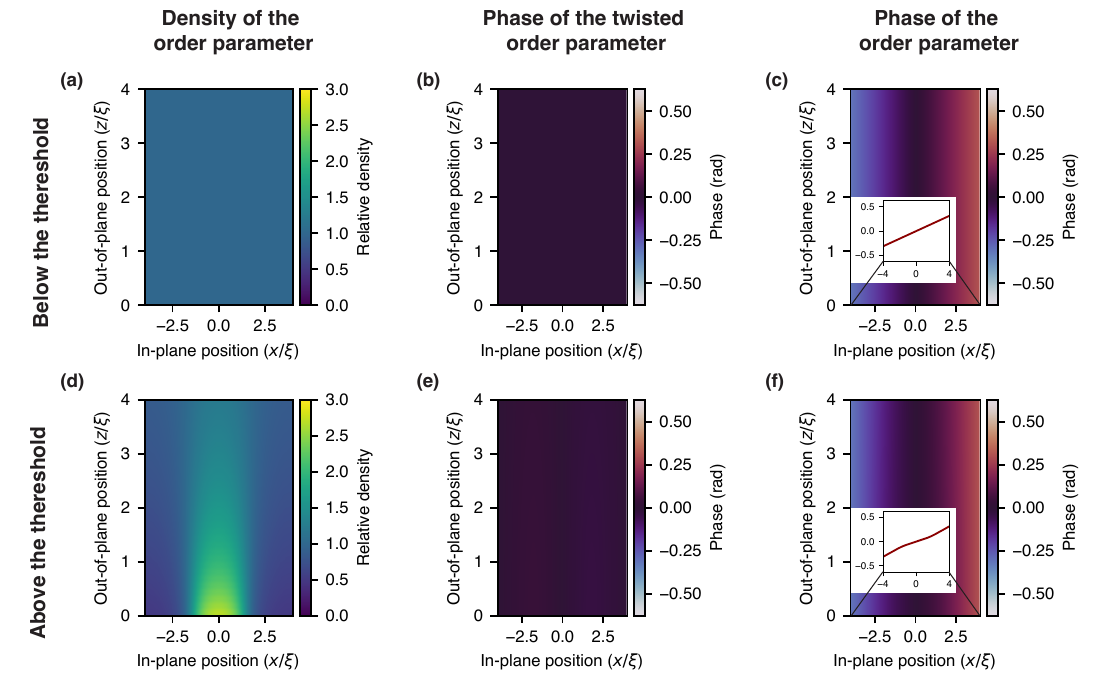}
\caption{\rev{\tb{Numerical results for the ground state in the presence of a phase twist.	} 
\tb{(a-c)} Ground state density and phase of the BEC for $\alpha=0.8\; \alpha_c$ and a phase twist $\mc{Q}=0.1\times 2\pi/L_x$ as a function of space. Here, $g=0.75$ meV.nm$^3$,  $\alpha_c\approx 0.16$ meV.nm$^3$,  $\xi\approx 6.3$ nm and $L_x=8\xi=50$ nm.
\tb{(d-f)} Ground state density and phase of the BEC for $\alpha=1.1\; \alpha_c$ as a function of space. Same phase twist and BEC parameters as above. 
\tb{(a)-(d)} Density $n/n_0$. 
\tb{(b)-(e)} Phase of the phase twisted order parameter $\tilde\psi$.
\tb{(c)-(f)} Phase of the untwisted order parameter $\psi$. The inset is the phase at the interface $z=0$. 
}}
\end{figure}

We discretize space in a box of size $L_x\times L_y \times L_z$ with periodic boundary conditions in $x$ and $y$. To recover bulk properties far from the interface we use $L_z=8\xi$ with 256 points. The system is invariant in $y$. We can vary $L_x$ and typically take $L_x=8\xi$ with 128 points. The result is robust on these parameters.

The ground state is obtained by solving the imaginary--time Gross--Pitaevskii equation:
\begin{equation}
  \partial_{\tau}  \psi(\r,\tau)
  =
\Delta \psi(\r,\tau) - (| \psi(\r,\tau)|^2-1)  \psi(\r,\tau) 
  \label{eq:imaginary_time_eq}
\end{equation}
with the Robin boundary condition at $z=0$. 
Far from the interface we expect to recover bulk properties. We therefore fix $\psi=1$ at $z=L_z$. 
We implement the Laplacian in the $z$-direction as well as the boundary conditions through a Crank--Nicolson scheme.
 The Laplacian in the $x$ and $y$ coordinates are implemented through Fourier space and the nonlinear terms is applied locally. 
 
 For strong interfacial couplings and in the absence of saturation for $B$ the order parameter is found to focus on one point of the interface. Indeed, the interfacial energy always dominates the volume energy cost if the order parameter can focus arbitrarily close to the interface. This collapse is prevented by implementing the saturation for $B$ (Eq. \eqref{B_sat}). In practice we fix 
\beq B_{\rm max} = \frac{2m}{\hbar^2} \frac{\alpha_c^2}{K_{\rm sol}} \eeq
which allows to observe the phase transition while preventing any collapse and we treat $B(r,t)$ as a dynamical variable evolving according to the energy functional and using a relaxation time of 1 (in dimensionless units).
 
We start from an initial guess which breaks translation symmetry for the first harmonics of the system:
 \beq \psi_0(\r) = 1+\left(\sum_{n=1}^N \epsilon_n\cos\left(n\frac{2\pi}{L_x}x\right)\right)e^{-z + i\left(\sum_{n=1}^N \epsilon_n\sin\left(n\frac{2\pi}{L_x}x\right)\right)} \eeq
 We typically take $N=4$ and $\epsilon_n$ ranging between 0.01 and 0.1. We further add some random noise which a typical amplitude of 1\% of the initial guess. The result is robust on these parameters.

We then apply imaginary time evolution with a timestep $\Delta \tau \leq 0.01$ for a total imaginary time $\tau_{\rm max}=20$. 
The result is robust on these parameters.
We check that energy decreases and saturates to a minimum while the order parameter converges.

Below the critical interfacial coupling constant $\alpha_c$ the order parameter converges towards the uniform bulk value $\psi(\r)=1$. Just above the threshold the first harmonics stabilizes while the higher ones vanish. The final state then breaks translation symmetry. 

We can further add a phase twist $\mc{Q}$ on the system in the $x$-direction. Thus, the phase twisted order parameter $\tilde \psi(\r,t) = \psi(\r,t)e^{i\mc{Q}x}$ solves the same Gross--Pitaevskii equation as before with a shifted $x$-derivative $\partial_x\rightarrow \partial_x + i\mc{Q}$. We then compute the ground state in the presence of a phase twist and starting with phase fluctuations. For a small but finite phase twist $\mc{Q}$ a superfluid phase should maintain the phase locked and $\tilde \psi$ should remain globally in-phase. In practice, we tested $\mc{Q}$ between $10^{-4}\times 2\pi/L_x$ and  $10^{-1}\times2\pi/L_x$ and found that the phase twisted order parameter remains globally in-phase over this range of parameters. Moreover, the density profile of the ground state was not impacted by the phase twist. Typical examples of density and phase of the ground state in the presence of a phase twist are given in Fig. S1. From these simulations we further extracted the superfluid phase stiffness, the superfluid density and the superfluid fraction.
